\documentclass[pre,showpacs,showkeys,preprintnumbers,amsmath,amssymb,superscriptaddress,twocolumn]{revtex4-2}
\usepackage{graphicx}
\usepackage{amsfonts}
\usepackage{url}

\usepackage{color}
\usepackage{epstopdf}
\usepackage{nameref,hyperref}
\usepackage{amsmath} 
\usepackage{color}
\usepackage{bm}

\def\x{\bm{x}}
\def\X{\bm{X}}

\def\E{\mathbb{E}}
\def\P{\mathbb{P}}
\def\R{\mathbb{R}}
\def\L{\mathcal{L}}
\def\M{\mathcal{M}}
\def\N{\mathcal{N}}
\def\T{\mathcal{T}}
\def\pa{\partial\Omega}

\def\ctanh{\mathrm{ctanh}}
\def\V{\mathbb{V}}

\begin{document}

\title{An encounter-based approach to the escape problem}

\author{Denis~S.~Grebenkov}
 \email{denis.grebenkov@polytechnique.edu}
\affiliation{
Laboratoire de Physique de la Mati\`{e}re Condens\'{e}e, \\ 
CNRS -- Ecole Polytechnique, Institut Polytechnique de Paris, 91120 Palaiseau, France}

\date{\today}

\begin{abstract}
We revise the encounter-based approach to imperfect
diffusion-controlled reactions, which employs the statistics of
encounters between a diffusing particle and the reactive region to
implement surface reactions.  We extend this approach to deal with a
more general setting, in which the reactive region is surrounded by a
reflecting boundary with an escape region.  We derive a spectral
expansion for the full propagator and investigate the behavior and
probabilistic interpretations of the associated probability flux
density.  In particular, we obtain the joint probability density of
the escape time and the number of encounters with the reactive region
before escape, and the probability density of the first-crossing time
of a prescribed number of encounters.  We briefly discuss
generalizations of the conventional Poissonian-type surface reaction
mechanism described by Robin boundary condition and potential
applications of this formalism in chemistry and biophysics.
\end{abstract}

\pacs{02.50.-r, 05.40.-a, 02.70.Rr, 05.10.Gg}



\keywords{Diffusion-controlled reaction, escape problem, first-passage time, surface reaction, 
boundary local time, encounters, Robin boundary condition, reflected Brownian motion, restricted diffusion}

\maketitle

\section{Introduction}
\label{sec:intro}

Diffusion-controlled reactions play an important role for various
chemical and biophysical applications
\cite{Alberts,Rice,Lauffenburger,Schuss,Lindenberg,Bressloff13}.
In a typical setting, a particle diffuses inside a confining domain
$\Omega$ toward a target region, on which it can react or trigger a
specific event.  One can think of a protein searching for a specific
site on a DNA molecule, or a molecule in a chemical reactor searching
for a catalytic germ to be transformed.  Such diffusion-controlled
reactions are often described in terms of the first-passage time to
the target or, more generally, to the reaction event
\cite{Redner,Metzler,Condamin07,Benichou10,Holcman13,Benichou14,Grebenkov16,Guerin16,Lanoiselee18}.
The diffusive dynamics of a single molecule is usually characterized
by a propagator, $G_q(\x,t|\x_0)$, i.e., the probability density that
a molecule started from $\x_0$ at time $0$ has arrived in a vicinity
of point $\x$ at a later time $t$, without being reacted
\cite{Redner,Gardiner,VanKampen}.  For ordinary diffusion, the
propagator satisfies the diffusion (or heat) equation,
\begin{equation}  \label{eq:Gq_diff}
\partial_t G_q(\x,t|\x_0) = D \Delta G_q(\x,t|\x_0)  \quad (\x\in \Omega),
\end{equation}
subject to the initial condition $G_q(\x,0|\x_0) = \delta(\x-\x_0)$
with a Dirac distribution $\delta(\x-\x_0)$, where $D$ is the
diffusion coefficient of the molecule, and $\Delta$ is the Laplace
operator (the meaning of the subscript $q$ will be explained below).
In turn, the reactive properties of the boundary of the confining
domain are usually incorporated through the Robin boundary condition,
\begin{equation}  \label{eq:Gq_Robin}
-D \partial_n G_q(\x,t|\x_0) = \kappa(\x) \, G_q(\x,t|\x_0)  \quad (\x\in\pa),
\end{equation}
where $\partial_n$ is the normal derivative on the boundary, oriented
outwards the confining domain, and $\kappa(\x)$ is the reactivity at a
boundary point $\x$.  This condition, which was put forward in
chemical physics by Collins and Kimball \cite{Collins49} and broadly
employed afterwards
\cite{Berg77,Sano79,Brownstein79,Weiss86,Powles92,Sapoval94,Sapoval02,Grebenkov05,Traytak07,Singer08,Bressloff08,Grebenkov10,Lawley15,Grebenkov15,Serov16,Bressloff17,Grebenkov17,Piazza19},
{\it postulates} that the diffusive flux from the bulk (the left-hand
side) is proportional to the propagator on the boundary.  Depending on
the type of surface reaction, the proportionality coefficient
$\kappa(\x)$ (in units m/s) is called reactivity, permeability, or
surface relaxivity.  In the context of bimolecular reactions, it can
also be related to the forward reaction constant.  The reactivity
$\kappa(\x)$ can range from $0$ for inert impermeable boundary to
$+\infty$ for a perfectly reactive boundary, on which the reaction
occurs upon the first arrival of the particle onto that boundary.

A space-dependent reactivity $\kappa(\x)$ allows one to implement
heterogeneous patterns on a catalytic surface or to describe in a
unified way the effects of reactive targets and restricting inert
boundaries of a porous medium.  However, theoretical description of
such general diffusion-controlled reactions is rather challenging (see
\cite{Grebenkov19b} and references therein).  For this reason, one
often focuses on a simpler yet relevant setting of a constant or
piecewise constant reactivity.  If $\kappa(\x)$ is constant, one can
employ standard spectral expansions over the eigenfunctions of the
Laplace operator \cite{Redner,Gardiner,VanKampen}.  Moreover, for
simple confining domains such as spheres or parallelepipeds, these
eigenfunctions are known exactly that facilitates the analysis of
diffusion-controlled reactions
\cite{Carslaw,Crank,Thambynayagam,Grebenkov13}.  Another common
setting is the case of a reactive target surrounded by a reflecting
boundary so that $\kappa(\x) = \kappa$ on the target surface, and
$\kappa(\x) = 0$ on the reflecting boundary.  In this case, one deals
with mixed Robin-Neumann (or Dirichlet-Neumann for $\kappa = \infty$)
boundary conditions.  When the target is small, one can employ
powerful asymptotic tools to approximate various quantities such as,
for instance, the mean reaction time on the target or the decay rate
of the survival probability
\cite{Mazya85,Ward93,Kolokolnikov05,Singer06a,Singer06b,Singer06c,Schuss07,Benichou08,Pillay10,Cheviakov10,Cheviakov11,Cheviakov12,Holcman14,Agranov18,Grebenkov19c}.
Moreover, if the target is interpreted as a ``hole'' in the otherwise
reflecting impenetrable boundary, one speaks about the escape or exit
problem.  As the reaction is understood here as an escape event, the
reaction time is called the escape or exit time.  Finally, if there
are many targets, their competition for the diffusing particle can be
characterized by splitting probabilities
\cite{Traytak92,Condamin06,Chevalier11,Galanti16,Grebenkov19f,Grebenkov20f,Klinger22}.

\begin{figure}[t!]
\begin{center}
\includegraphics[width=50mm]{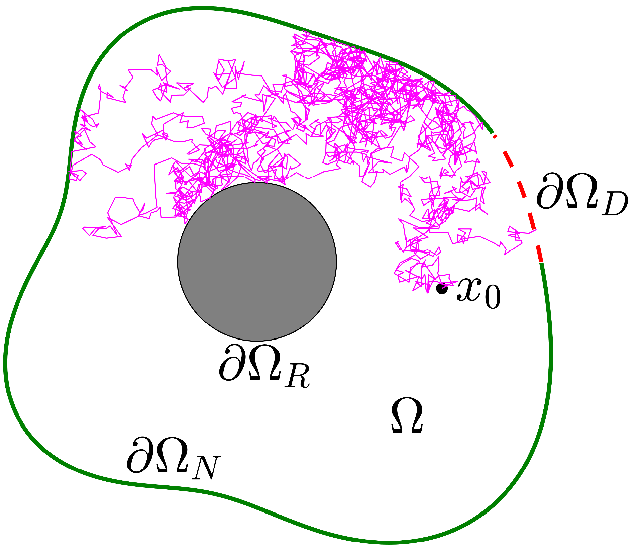} 
\end{center}
\caption{
Illustration of a confining domain $\Omega$ whose boundary $\pa$ is
split into three disjoint parts: a target region $\pa_R$ (the boundary
of an obstacle shadowed in gray, e.g., a catalytic germ), a reflecting
region $\pa_N$ (solid green line), and an escape region $\pa_D$
(dashed red line).  A simulated trajectory of a particle that started
from a point $\x_0$ and diffused until its escape, is shown in
magenta.  The particle is always reflected from $\pa_N$ but may either
be reflected from or react on $\pa_R$.}
\label{fig:domain}
\end{figure}

In this paper, we consider a more general situation, in which the
boundary $\pa$ of the confining domain is split into three disjoint
parts,
\begin{equation}  \label{eq:pa}
\pa = \pa_R \cup \pa_N \cup \pa_D,
\end{equation}
which represent the reactive target $\pa_R$, the inert reflecting
boundary $\pa_N$ and the escape region $\pa_D$
(Fig. \ref{fig:domain}).  In this way, one can describe an important
class of diffusion-controlled reactions, in which the diffusing
particle can leave the confining domain through an escape region or be
destroyed on it, without being reacted on the target region.  This is
a common setting for many biochemical reactions inside a living cell;
for instance, proteins can be disassembled before finding their
receptors, while ions can leave the cytoplasm through the ion channels
on the plasma membrane.  We assume that the particle disappears
immediately after the first arrival on the escape region $\pa_D$.  In
mathematical terms, such a composed boundary can be implemented
through the mixed Robin-Dirichlet-Neumann boundary conditions:
\begin{subequations}  \label{eq:Gq_BC}
\begin{align}  \label{eq:Gq_R}
\partial_n G_q(\x,t|\x_0) + q G_q(\x,t|\x_0) & = 0 \quad (\x\in\pa_R), \\
G_q(\x,t|\x_0) & = 0 \quad (\x\in\pa_D), \\  \label{eq:Gq_N}
\partial_n G_q(\x,t|\x_0) & = 0 \quad (\x\in\pa_N) ,
\end{align}
\end{subequations}
where $q = \kappa/D$ is proportional to the (constant) reactivity
$\kappa$ of the target region $\pa_R$ (note that the subscript of
$\pa_R$, $\pa_D$ and $\pa_N$ refers to the corresponding type of
Robin, Dirichlet and Neumann boundary condition).  Alternatively,
these conditions can describe a two-target problem: a partially
reactive target $\pa_R$ and a perfectly reactive target $\pa_D$
(``reaction'' on $\pa_D$ is interpreted here as the escape event).
However, as we focus on reactions on the target region $\pa_R$, we
keep speaking about a single-target problem in the presence of escape
events.

In order to solve this problem, one could still introduce the
Laplacian eigenfunctions that satisfy the same mixed boundary
conditions.  Here we follow an alternative way and generalize the
encounter-based approach that was developed in \cite{Grebenkov20} for
the particular case of a constant reactivity on the whole boundary
(when $\pa = \pa_R$ and $\pa_N = \pa_D = \emptyset$).  This approach
relies on the concept of the boundary local time $\ell_t$, which can
represent a rescaled number of encounters between the diffusing
particle and the boundary up to time $t$.  In this way, one can first
investigate the diffusive dynamics inside a confining domain with
reflecting boundary and then implement surface reactions explicitly
(see below).  Moreover, one can go beyond the conventional
Poissonian-type reactions described by Robin boundary condition
(\ref{eq:Gq_R}) and implement saturation or activation effects or,
more generally, encounter-dependent reactivity \cite{Grebenkov20}.
Our generalization allows one to investigate the effects of the escape
event onto such reactions.  We also provide probabilistic
interpretations of the probability flux density that were not
discussed enough in earlier works.  In particular, we obtain the joint
probability density of the position, boundary local time and the
escape time.

The paper is organized as follows.  In Sec. \ref{sec:main}, we present
the main theoretical results.  After introducing the necessary
elements of the conventional and encounter-based approaches in
Sec. \ref{sec:conventional} and Sec. \ref{sec:encounter}, we discuss
the distribution of the boundary local time (Sec. \ref{sec:rho}) and
restrictions of the probability flux density to $\pa_D$ and $\pa_R$
(Secs. \ref{sec:jD} and \ref{sec:jR}).  In Sec. \ref{sec:sphere}, we
illustrate the behavior of the derived quantities for a particle
diffusing between two concentric spheres.  In this case, all the
``ingredients'' of the encounter-based approach can be found
explicitly.  Section \ref{sec:discussion} presents further
discussions, conclusions and perspectives.

\section{Main results}
\label{sec:main}

\subsection{Conventional approach}
\label{sec:conventional}

In order to highlight the advantages of the encounter-based approach,
we briefly recall several characteristics of diffusion-controlled
reactions accessed within the conventional approach that relies on the
propagator $G_q(\x,t|\x_0)$.  We consider a point-like particle
diffusing in a bounded Euclidean domain $\Omega\subset \R^d$ with
smooth boundary $\pa$, which is partitioned into reactive ($\pa_R$),
reflecting ($\pa_N$) and escape ($\pa_D$) parts.  As the diffusing
particle can disappear due to either escape through $\pa_D$ or
reaction on $\pa_R$, one can naturally introduce two first-passage
times: the escape time
\begin{equation}  \label{eq:Tdef}
T_q = \inf\{ t>0 ~:~ \X_t \in \pa_D\}
\end{equation}
as the first-passage time to the escape region $\pa_D$, and the
reaction time $\tau_q$ as the random instance of the reaction event on
$\pa_R$ (its formal definition is nontrivial and will be given in
Sec. \ref{sec:jR}).  The subscript $q$ highlights that the
distributions of both random variables $T_q$ and $\tau_q$ depend on
the reactivity of the target region because they are determined by the
probability flux density,
\begin{equation}  \label{eq:jq_def}
j_q(\x,t|\x_0) = - D\partial_n G_q(\x,t|\x_0)  \quad (\x\in \pa).
\end{equation}
In fact, the restriction of $j_q(\x,t|\x_0)$ to the escape region
$\pa_D$ is the joint probability density of the escape location
$\X_{T_q}$ and its time $T_q$.  In turn, the restriction of
$j_q(\x,t|\x_0)$ to the target region $\pa_R$ is the joint probability
density of the reaction location $\X_{\tau_q}$ and its time $\tau_q$
(note that the restriction of $j_q(\x,t|\x_0)$ to $\pa_N$ is strictly
zero due to the Neumann boundary condition (\ref{eq:Gq_N})).  If the
position does not matter, it can be averaged out to get the (marginal)
probability densities of the escape time $T_q$ and of the reaction
time $\tau_q$:
\begin{subequations}
\begin{align}  \label{eq:JqD_def}
J_q^D(t|\x_0) &= \int\limits_{\pa_D} d\x \, j_q(\x,t|\x_0) , \\  \label{eq:JqR_def}
J_q^R(t|\x_0) &= \int\limits_{\pa_R} d\x \, j_q(\x,t|\x_0) .
\end{align}
\end{subequations}
As the particle can either escape or react, none of these densities is
normalized to $1$; in turn, one has
\begin{equation}  \label{eq:JqDR_norm}
\int\limits_0^\infty dt \, \bigl[J_q^D(t|\x_0) + J_q^R(t|\x_0)\bigr] = 1.
\end{equation}
To show this normalization, one can integrate the diffusion equation
(\ref{eq:Gq_diff}) over $\x\in \Omega$ and use the Green's formula and
mixed boundary conditions (\ref{eq:Gq_BC}) to get the continuity
equation
\begin{equation}  \label{eq:Sq_cont}
\partial_t S_q(t|\x_0) = - \bigl[J_q^D(t|\x_0) + J_q^R(t|\x_0)\bigr],
\end{equation}
where
\begin{equation}
S_q(t|\x_0) = \int\limits_{\Omega} d\x \, G_q(\x,t|\x_0)
\end{equation}
is the survival probability of the particle up to time $t$.  The
integral of Eq. (\ref{eq:Sq_cont}) over $t$ from $0$ to infinity
yields the normalization (\ref{eq:JqDR_norm}), given that $S_q(0|\x_0)
= 1$ and $S_q(t|\x_0) \to 0$ as $t\to\infty$ for diffusion in a
bounded domain.

\subsection{Encounter-based approach}
\label{sec:encounter}

In the encounter-based approach, one first characterizes purely
diffusive dynamics inside a bounded confining domain $\Omega$ with a
{\it reflecting} boundary $\pa$, and then incorporates surface
reactions on $\pa$, i.e., transforms the reflecting boundary into the
reactive one \cite{Grebenkov20}.  For this purpose, one uses the
boundary local time $\ell_t$, which was first introduced by L\'evy
\cite{Levy} and then extensively employed in mathematical literature
on stochastic processes \cite{Ito,Freidlin}.  The boundary local time
can be defined as the renormalized residence time near the boundary
$\pa$:
\begin{equation}  \label{eq:ellt_def}
\ell_t = \lim\limits_{a\to 0} \frac{D}{a} \int\limits_0^t dt' \, \Theta(a - |\X_{t'} - \pa|),
\end{equation}
where $|\X_{t'} - \pa|$ is the Euclidean distance between the position
$\X_{t'}$ of the particle at time $t'$ and the boundary $\pa$, and
$\Theta(z)$ is the Heaviside step function: $\Theta(z) = 1$ for $z>0$
and $0$ otherwise.  The integral in Eq. (\ref{eq:ellt_def}) defines
the residence (also known as occupation or sojourn) time in a thin
layer of width $a$ near the boundary.  As the boundary $\pa$ has a
lower dimension as compared to the domain $\Omega$, the residence time
vanishes as the layer shrinks (when $a\to 0$); in turn, its
renormalization by $a$ yields the nontrivial limit
(\ref{eq:ellt_def}).  The boundary local time should not be confused
with the local time in a bulk point, which has been intensively
studied, especially for diffusive processes in one dimension
\cite{Borodin,Majumdar05}.  Despite its name, the boundary local time has
units of length.  The boundary local time can be equivalently defined
as a rescaled limit of the number $\N_t^a$ of downcrossings of the
boundary layer of width $a$ up to time $t$: $\ell_t =
\lim\limits_{a\to 0} a \N_t^a$.  As each downcrossing can be
interpreted as an encounter of the particle with the boundary, the
boundary local time $\ell_t$ characterizes the statistics of such
encounters \cite{Grebenkov21}.  The diffusive dynamics can then be
described either by a stochastic differential equation for the random
pair $(\X_t,\ell_t)$, or by the so-called full propagator
$P(\x,\ell,t|\x_0)$, i.e., the joint probability density of getting
the values $(\x,\ell)$ for the pair $(\X_t,\ell_t)$.  Once the full
propagator is known, one can implement various surface reaction
mechanisms (see Sec. \ref{sec:jR}).

By construction, the boundary local time $\ell_t$ characterizes
encounters with the whole boundary $\pa$ that does not allow one to
implement different reaction mechanisms on different subsets of the
boundary.  This problem has been discussed and partly resolved in
\cite{Grebenkov20c} by introducing a proper boundary local time
$\ell_t^i$ on each subset of interest.  In this case, one would deal
with the joint probability density for $\X_t$ and for boundary local
times on all these subsets.  Even though a formal way for computing
this density was proposed in \cite{Grebenkov20c}, its practical
implementation was realized only for simple geometric settings (e.g.,
an interval).

In the context of the escape problem that we consider in this paper,
one can follow a different strategy.  As we are interested in
describing reactions exclusively on the target region $\pa_R$, we need
to know the statistics of encounters with that particular region.  In
other words, one can modify the above definition of the boundary local
time to count encounters only with $\pa_R$:
\begin{equation}  \label{eq:elltR_def}  
\ell_t^R = \lim\limits_{a\to 0} \frac{D}{a} \int\limits_0^t dt' \, \Theta(a - |\X_{t'} - \pa_R|).
\end{equation}
This relation defines a non-decreasing stochastic process $\ell_t^R$
(starting from $\ell_0^R = 0$) that increases at each encounter with
$\pa_R$.  As previously, the integral is the residence time that a
particle spent in a thin layer of width $a$ near the target region
$\pa_R$ up to time $t$.  When $a$ is small, this residence time can
thus be approximated as $\ell_t^R a/D$ according to
Eq. (\ref{eq:elltR_def}).  In the same vein, we introduce the full
propagator $P(\x,\ell,t|\x_0)$ as the joint probability density for
$\X_t$ and $\ell_t^R$ (not $\ell_t$), under the condition of no escape
through $\pa_D$ up to time $t$.  As previously, the target region
$\pa_R$ is treated at this step as {\it reflecting}, i.e., the
particle described by the full propagator can disappear only on the
escape region $\pa_D$.

At the next step, one can introduce reaction events on $\pa_R$
following the probabilistic arguments from \cite{Grebenkov20}.  For
this purpose, one can consider a thin reactive layer of width $a$ near
the target region $\pa_R$.  Once the particle enters this layer,
surface reaction may be described by a standard first-order reaction
kinetics, with the rate $k = \kappa/a$.  Since the residence time of
the particle within this layer up to time $t$ is approximately equal
to $\ell_t^R a/D$, the probability of no reaction on $\pa_R$ up to $t$
is then $e^{-k(\ell_t^R a/D)} = e^{-q\ell_t^R}$, where $q = \kappa/D$.
As a consequence, one deduces the following relation between the
conventional and full propagators:
\begin{align}  \nonumber 
G_q(\x,t|\x_0) &  = \E_{\x_0} \{ \delta(\X_t - \x) e^{-q\ell_t^R} \Theta(T_q - t)\} \\   \label{eq:Gq_P} 
& = \int\limits_0^\infty d\ell \, e^{-q\ell} \, P(\x,\ell,t|\x_0).
\end{align}
On the left-hand side, the conventional propagator $G_q(\x,t|\x_0)$
satisfying Eqs. (\ref{eq:Gq_diff}, \ref{eq:Gq_BC}), describes
diffusion from $\x_0$ to $\x$ in time $t$, under the condition of no
escape through $\pa_D$ and no reaction on $\pa_R$ up to $t$.  The
probability density of this event can be written via the expectation
in the middle, i.e. as the fraction of trajectories $\X_t$ of
reflected Brownian motion between $\x_0$ and $\x$ of duration $t$,
with the penalizing factor $e^{-q\ell_t^R}$ eliminating a subset of
trajectories that reacted on $\pa_R$, while $\Theta(T_q - t)$
eliminating those that have escaped before $t$.  On the right-hand
side, the full propagator $P(\x,\ell,t|\x_0)$ describes diffusion from
$\x_0$ to $\x$ in time $t$, under the condition of no escape through
$\pa_D$ and of getting the boundary local time $\ell_t^R$ equal to
$\ell$.  In turn, the factor $e^{-q\ell}$ incorporates the probability
of no reaction on $\pa_R$ for any realized value $\ell$ of the
boundary local time $\ell_t^R$, while the integral over $\ell$ sums up
contributions from all possible realizations of $\ell_t^R$.  In other
words, this integral simply evaluates the expectation in the middle.
We stress that the condition $T_q > t$ of no escape through $\pa_D$ is
implemented in the full propagator $P(\x,\ell,t|\x_0)$ through the
Dirichlet boundary condition on $\pa_D$ (see below).

The relation (\ref{eq:Gq_P}) plays the central role in this work.  On
the one hand, it allows one to incorporate the conventional surface
reactions as the Laplace transform of the full propagator
$P(\x,\ell,t|\x_0)$ with respect to $\ell$.  Importantly, the
reactivity parameter $q$ enters {\it explicitly} through the
exponential factor $e^{-q\ell}$, whereas it appeared {\it implicitly}
in the conventional description as a coefficient in Robin boundary
condition (\ref{eq:Gq_R}).  Finally, one can replace the factor
$e^{-q\ell}$, which is reminiscent of an exponential probability law,
by another function, allowing one to implement various surface
reaction mechanisms (see Sec. \ref{sec:jR}).  On the other hand, the
inverse Laplace transform of the conventional propagator
$G_q(\x,t|\x_0)$ with respect to $q$ gives access to the full
propagator:
\begin{equation}  \label{eq:P_inv_Gq}
P(\x,\ell,t|\x_0) = \L_{q\to\ell}^{-1} \bigl\{ G_q(\x,t|\x_0)\bigr\} .
\end{equation}
Unfortunately, an implicit dependence of $G_q(\x,t|\x_0)$ on $q$ often
prevents using this inversion and thus urges for another
representation for the full propagator.

To achieve this goal, we extend the spectral expansion of the full
propagator developed in \cite{Grebenkov20}.  For this purpose, we
introduce an extension of the so-called Dirichlet-to-Neumann operator
$\M_p$ that associates to a given function $f$ on the target region
$\pa_R$ another function $g$ on that region such that 
\begin{equation}
\M_p f = g = \left. (\partial_n u)\right|_{\pa_R} ,
\end{equation}
where $u$ is the solution of the following boundary value problem with
$p \geq 0$:
\begin{subequations}
\begin{align}
(p - D\Delta) u & = 0  \quad (\x\in\Omega) , \\
u & = f \quad (\x\in\pa_R), \\
u & = 0 \quad (\x\in \pa_D) , \\
\partial_n u & = 0 \quad (\x\in \pa_N).
\end{align}
\end{subequations}
In other words, the operator $\M_p$ transforms the Dirichlet boundary
condition $u = f$ on $\pa_R$ into an equivalent Neumann boundary
condition $\partial_n u = g$ on $\pa_R$, keeping unchanged the
Dirichlet and Neumann conditions on $\pa_D$ and $\pa_N$ respectively.
While the conventional Dirichlet-to-Neumann operator acted on
functions on the whole boundary $\pa$, our extension acts on functions
on the subset $\pa_R$ of the boundary.  The spectral properties of the
conventional Dirichlet-to-Neumann operator have been intensively
studied in mathematical literature
\cite{Arendt14,Daners14,terElst14,Behrndt15,Arendt15,Hassell17,Girouard17}.
Most of these properties are expected to be valid for our extension so
that $\M_p$ is a pseudo-differential self-adjoint operator.  Since the
target region $\pa_R$ is bounded, the spectrum of $\M_p$ is discrete,
with an infinite set of positive eigenvalues $\mu_k^{(p)}$ ($k =
0,1,2,\ldots$), that can be enumerated in the increasing order:
\begin{equation}
0 \leq \mu_0^{(p)} \leq \mu_1^{(p)} \leq \ldots \leq \mu_k^{(p)} \leq \ldots \nearrow +\infty .
\end{equation}
In turn, the associated eigenfunctions $v_k^{(p)}(\x)$ form a complete
orthonormal basis of the functional space $L_2(\pa_R)$ of square
integrable functions on $\pa_R$.  A rigorous formulation and
demonstration of these mathematical properties are beyond the scope of
this paper.  From the mathematical point of view, one can consider
them as conjectural extensions of the well-known conventional case.

The eigenbasis of the Dirichlet-to-Neumann operator can serve for
getting a spectral expansion of the full propagator.  Skipping
technical details (given in \cite{Grebenkov20}), we sketch here the
main steps of this derivation.  In the first step, the Laplace
transform of the conventional propagator $G_q(\x,t|\x_0)$ with respect
to time $t$,
\begin{equation}
\tilde{G}_q(\x,p|\x_0) = \int\limits_0^\infty dt \, e^{-pt} \, G_q(\x,t|\x_0),
\end{equation}
reduces the diffusion equation (\ref{eq:Gq_diff}) to the inhomogeneous
modified Helmholtz equation:
\begin{equation}  \label{eq:Gq_Helm}
(p - D\Delta) \tilde{G}_q(\x,p|\x_0) = \delta(\x-\x_0),
\end{equation}
subject to the same boundary condition (here and below, tilde denotes
Laplace-transformed quantities with respect to $t$).  Writing
$\tilde{G}_q(\x,p|\x_0) = \tilde{G}_\infty(\x,p|\x_0) +
\tilde{g}_q(\x,p|\x_0)$ with an unknown function $\tilde{g}_q(\x,p|\x_0)$,
one eliminates $\delta(\x-\x_0)$ from the right-hand side.  As
$\tilde{g}_q(\x,p|\x_0)$ satisfies $(p - D\Delta)
\tilde{g}_q(\x,p|\x_0) = 0$, one can employ the eigenfunctions of $\M_p$
for a spectral decomposition of the restriction of
$\tilde{g}_q(\x,p|\x_0)$ to $\pa_R$, which can then be extended to the
whole domain $\Omega$.  Finally, one takes the inverse Laplace
transform of $\tilde{G}_q(\x,p|\x_0)$ with respect to $q$ to get the
spectral expansion of the full propagator in the Laplace domain:
\begin{align}  \label{eq:P_spectral}
\tilde{P}(\x,\ell,p|\x_0) & = \tilde{G}_\infty(\x,p|\x_0) \delta(\ell) \\ \nonumber
& + \frac{1}{D} \sum\limits_{k=0}^\infty [V_k^{(p)}(\x_0)]^* \, V_k^{(p)}(\x)\,  e^{-\mu_k^{(p)}\ell} ,
\end{align}
where asterisk denotes complex conjugate, and
\begin{equation}  \label{eq:Vk}
V_k^{(p)}(\x_0) = \int\limits_{\pa_R} d\x \, \tilde{j}_\infty(\x,p|\x_0) \, v_k^{(p)}(\x)
\end{equation}
is the extension of the eigenfunction $v_k^{(p)}(\x)$ (defined on
$\pa_R$) to the whole domain $\Omega$.  While the structure of the
spectral expansion (\ref{eq:P_spectral}) is identical to that derived
in \cite{Grebenkov20}, its ``ingredients''
$\tilde{G}_\infty(\x,p|\x_0)$, $V_k^{(p)}(\x_0)$ and $\mu_k^{(p)}$ are
adapted to account for the presence of reflecting and escape regions.

One can easily check that the functions $V_k^{(p)}(\x)$ defined by
Eq. (\ref{eq:Vk}) satisfy:
\begin{subequations}  \label{eq:Vk_problem}
\begin{align}  \label{eq:Vk_Helm}
(p - D \Delta) V_k^{(p)}(\x) &= 0 \quad (\x\in\Omega), \\
V_k^{(p)}(\x) & = v_k^{(p)}(\x) \quad (\x\in \pa_R), \\  \label{eq:Vk_D}
V_k^{(p)}(\x) & = 0 \quad (\x\in \pa_D), \\  \label{eq:Vk_N}
\partial_n V_k^{(p)}(\x) & = 0 \quad (\x\in \pa_N).
\end{align}
\end{subequations}
Moreover, since $v_k^{(p)}$ is an eigenfunction of the
Dirichlet-to-Neumann operator, one has
\begin{equation}  \label{eq:dVk_R}
\partial_n V_k^{(p)}(\x) = \mu_k^{(p)} V_k^{(p)}(\x) = \mu_k^{(p)} v_k^{(p)}(\x)  \quad (\x\in\pa_R).
\end{equation}
Once the eigenfunction $v_k^{(p)}(\x)$ is found, one can determine its
extension $V_k^{(p)}(\x)$ either via Eq. (\ref{eq:Vk}), or by solving
the above problem (\ref{eq:Vk_problem}).  Alternatively, without
knowing $v_k^{(p)}(\x)$, one can look directly at the eigenvalue
problem (\ref{eq:Vk_Helm}, \ref{eq:Vk_D}, \ref{eq:Vk_N},
\ref{eq:dVk_R}), in which the spectral parameter (here, $\mu_k^{(p)}$)
stands in the boundary condition.  This is known as the Steklov
problem (see \cite{Girouard17} and references therein), while
$\mu_k^{(p)}$ and $V_k^{(p)}(\x)$ are the eigenvalues and
eigenfunctions of this problem.  Despite this equivalence, we keep
referring to the Dirichlet-to-Neumann operator $\M_p$ and its spectral
properties.

The inverse Laplace transform of Eq. (\ref{eq:P_spectral}) with
respect to $p$ formally yields
\begin{align}  \label{eq:Pt_spectral}
P(\x,\ell,t|\x_0) & = G_\infty(\x,t|\x_0) \delta(\ell) \\ \nonumber
& + \L_{p\to t}^{-1} \biggl\{\frac{1}{D} \sum\limits_{k=0}^\infty 
[V_k^{(p)}(\x_0)]^* \, V_k^{(p)}(\x)\,  e^{-\mu_k^{(p)}\ell}  \biggr\}.
\end{align}
The first term in Eq. (\ref{eq:Pt_spectral}) represents the
contribution of random trajectories from $\x_0$ to $\x$ of duration
$t$ without hitting neither the target region $\pa_R$, nor the escape
region $\pa_D$ (in turn, they could encounter the reflecting part
$\pa_N$).  Their ``fraction'' is precisely given by
$G_\infty(\x,t|\x_0)$, with Dirichlet boundary condition on $\pa_D$
and $\pa_R$ (note that Eq. (\ref{eq:Gq_R}) becomes
$G_\infty(\x,t|\x_0) = 0$ for $q = \infty$).  As the boundary local
time $\ell_t^R$ remained zero for these trajectories, one gets the
singular factor $\delta(\ell)$.  In turn, the second term accounts for
all trajectories that have encountered the target region $\pa_R$, but
still avoided the escape through $\pa_D$ (the latter condition is
implemented via Eq. (\ref{eq:Vk_D}) for all $V_k^{(p)}(\x)$).  The
spectral representation (\ref{eq:Pt_spectral}) is an alternative way
for computing the full propagator, which is complementary to
Eq. (\ref{eq:P_inv_Gq}).  While both expressions involve an inverse
Laplace transform, the spectral characteristics of the
Dirichlet-to-Neumann operator are in general easier to access than the
conventional propagator $G_q(\x,t|\x_0)$ in time domain.  For
instance, we will employ Eq. (\ref{eq:Pt_spectral}) in
Sec. \ref{sec:sphere} to deduce various properties of
diffusion-controlled reactions in a spherical domain.  We emphasize
that two representations provide complementary insights onto the full
propagator.

The full propagator determines the corresponding probability flux
density on the boundary $\pa$:
\begin{equation}
j(\x,\ell,t|\x_0) = -D \partial_n P(\x,\ell,t|\x_0) \quad (\x\in\pa).
\end{equation}
The spectral expansion (\ref{eq:P_spectral}) gives access to this
quantity in the Laplace domain
\begin{align} \nonumber
\tilde{j}(\x,\ell,p|\x_0) & = \tilde{j}_\infty(\x,p|\x_0) \delta(\ell) \\   \label{eq:j_spectral}
& - \sum\limits_{k=0}^\infty [V_k^{(p)}(\x_0)]^* \, (\partial_n V_k^{(p)})(\x)\,  e^{-\mu_k^{(p)}\ell} .
\end{align}
The probability flux density $j(\x,\ell,t|\x_0)$ is the main object of
our study.  In particular, we aim at providing its probabilistic
interpretation and deducing various related characteristics of
diffusion-controlled reactions in the presence of escape events.  We
will show that the interpretation of $j(\x,\ell,t|\x_0)$ is more
subtle than that of $j_q(\x,t|\x_0)$ mentioned in
Sec. \ref{sec:conventional}.  For instance, it is easy to prove that
\begin{equation}  \label{eq:jR_int}
\int\limits_0^\infty d\ell \, j(\x,\ell,t|\x_0) = 0 \quad (\x\in\pa_R),
\end{equation}
i.e., the restriction of $j(\x,\ell,t|\x_0)$ to $\pa_R$ is not
necessarily positive.

To show this relation, let us first integrate
Eq. (\ref{eq:P_spectral}) over $\ell$ from $0$ to infinity to get the
following identity
\begin{align}  \label{eq:G0_Ginf}
\tilde{G}_0(\x,p|\x_0) & = \tilde{G}_\infty(\x,p|\x_0)  \\ \nonumber
& + \frac{1}{D} \sum\limits_{k=0}^\infty [V_k^{(p)}(\x_0)]^* \, V_k^{(p)}(\x) \frac{1}{\mu_k^{(p)}} \,,
\end{align}
so that
\begin{align}  \label{eq:j0_jinf}
\tilde{j}_0(\x,p|\x_0) & = \tilde{j}_\infty(\x,p|\x_0)  \\ \nonumber
& - \sum\limits_{k=0}^\infty [V_k^{(p)}(\x_0)]^* \, (\partial_n V_k^{(p)})(\x) \frac{1}{\mu_k^{(p)}} \,.
\end{align}
For any $\x \in \pa_R$, the left-hand side of Eq. (\ref{eq:j0_jinf})
is zero, implying
\begin{equation}  \label{eq:jinf_R}
\tilde{j}_\infty(\x,p|\x_0) = \sum\limits_{k=0}^\infty [V_k^{(p)}(\x_0)]^* \, v_k^{(p)}(\x)  \quad (\x\in\pa_R),
\end{equation}
where we applied Eq. (\ref{eq:dVk_R}).  Using the identity
(\ref{eq:jinf_R}), one can easily check that the integral of
Eq. (\ref{eq:j_spectral}) over $\ell$ from $0$ to infinity is strictly
zero for any $\x\in\pa_R$, that reads in time domain as
Eq. (\ref{eq:jR_int}).

In order to clarify the probabilistic meaning of $j(\x,\ell,t|\x_0)$,
we first look at the distribution of the boundary local time
$\ell_t^R$, then discuss the restriction of $j(\x,\ell,t|\x_0)$ to the
escape region $\pa_D$, and finally describe its restriction to the
target region $\pa_R$.

\subsection{Probability density of the boundary local time}
\label{sec:rho}

It is convenient to start by inspecting the distribution of the
boundary local time $\ell_t^R$.  By definition, the integral of the
full propagator $P(\x,\ell,t|\x_0)$ over $\x\in\Omega$ determines the
(marginal) probability density of the boundary local time
\begin{equation}
\rho(\ell,t|\x_0) = \int\limits_\Omega d\x \, P(\x,\ell,t|\x_0).
\end{equation}
Since the particle may escape the domain, this probability density is
not normalized to $1$:
\begin{equation}  \label{eq:rho_norm}
\int\limits_0^\infty d\ell \, \rho(\ell,t|\x_0) = S_0(t|\x_0) = \P_{\x_0}\{ T_0 > t\} ,
\end{equation}
i.e., the probability of no escape up to time $t$ (the subscript $q =
0$ of $T_0$ highlights that the target region $\pa_R$ is treated here
as reflecting).  Only if there is no escape region, the survival
probability $S_0(t|\x_0)$ is equal to $1$, ensuring the normalization
in the conventional case \cite{Grebenkov19a}.

Integrating the fundamental relation (\ref{eq:Gq_P}) over
$\x\in\Omega$, one gets 
\begin{equation}
S_q(t|\x_0) = \int\limits_0^\infty d\ell \, e^{-q\ell} \, \rho(\ell,t|\x_0) = \E_{\x_0}\{ e^{-q\ell_t^R}\} ,
\end{equation}
i.e., the survival probability is the generating function of
$\ell_t^R$.  Moreover, the inverse Laplace transform of
Eq. (\ref{eq:Sq_cont}) with respect to $q$ yields another continuity
equation
\begin{equation}  \label{eq:rho_cont}
\partial_t \rho(\ell,t|\x_0) = - \bigl[J_R(\ell,t|\x_0) + J_D(\ell,t|\x_0)\bigr],
\end{equation}
where
\begin{subequations}
\begin{align}  \label{eq:JR}
J_R(\ell,t|\x_0) &= \int\limits_{\pa_R} d\x \, j(\x,\ell,t|\x_0) , \\  \label{eq:JD}
J_D(\ell,t|\x_0) &= \int\limits_{\pa_D} d\x \, j(\x,\ell,t|\x_0) ,
\end{align}
\end{subequations}
and we used that
\begin{equation} \label{eq:jq_j}
j_q(\x,t|\x_0) = \int\limits_0^\infty d\ell \, e^{-q\ell} \, j(\x,\ell,t|\x_0)
\end{equation}
due to Eq. (\ref{eq:Gq_P}).

In the Laplace domain, the integral of Eq. (\ref{eq:P_spectral}) yields
\begin{align}  \label{eq:rho_spectral}
\tilde{\rho}(\ell,p|\x_0) & = \tilde{S}_\infty(p|\x_0) \delta(\ell) \\ \nonumber
& + \frac{1}{D} \sum\limits_{k=0}^\infty [V_k^{(p)}(\x_0)]^* \,  e^{-\mu_k^{(p)}\ell} 
\int\limits_{\Omega} d\x \, V_k^{(p)}(\x).
\end{align}

\subsection{Escape events}
\label{sec:jD}

In a direct analogy with $j_q(\x,t|\x_0)$, the restriction of the
probability flux density $j(\x,\ell,t|\x_0)$ to the escape region
$\pa_D$ determines the joint probability density of the position
$\X_{T_0}$, the boundary local time $\ell_{T_0}^R$, and the escape
time $T_0$ defined by Eq. (\ref{eq:Tdef}).
If the exact location of the escape does not matter, one can integrate
over $\x\in\pa_D$ to get the joint probability density
$J_D(\ell,t|\x_0)$ of $\ell_{T_0}^R$ and $T_0$, see Eq. (\ref{eq:JD}).
In the Laplace domain, this quantity reads
\begin{align}  \label{eq:JD_tilde}
\tilde{J}_D(\ell,p|\x_0) & = \tilde{J}_\infty^D(p|\x_0) \delta(\ell) \\  \nonumber
& + \sum\limits_{k=0}^\infty [V_k^{(p)}(\x_0)]^* \, e^{-\mu_k^{(p)}\ell}  C_k^{(p)},
\end{align}
where
\begin{equation}  \label{eq:Ck_def}
C_k^{(p)} = - \int\limits_{\pa_D} d\x\, \partial_n V_k^{(p)}(\x)
\end{equation}
and $\tilde{J}_\infty^D(p|\x_0)$ is given by Eq. (\ref{eq:JqD_def}) at
$q = \infty$.  While the escape time $T_0$ has been studied in the
past, the joint distribution of $\ell_{T_0}^R$ and $T_0$ has not been
reported earlier.

Integrating Eq. (\ref{eq:Vk_Helm}), using the Green's formula and
expressions (\ref{eq:Vk_N}, \ref{eq:dVk_R}), one gets another
representation:
\begin{equation}
C_k^{(p)} = \mu_k^{(p)} \int\limits_{\pa_R} d\x\, v_k^{(p)}(\x) - \frac{p}{D} \int\limits_{\Omega} d\x \, V_k^{(p)}(\x).
\end{equation}
One can apply this representation to check the correct normalization
of $J_D(\ell,t|\x_0)$:
\begin{align*}
& \int\limits_0^\infty d\ell \int\limits_0^\infty dt\, J_D(\ell,t|\x_0) 
= \int\limits_0^\infty d\ell \, \tilde{J}_D(\ell,p|\x_0) \\
& = \tilde{J}_\infty^D(0|\x_0) + \sum\limits_{k=0}^\infty [V_k^{(0)}(\x_0)]^* \frac{C_k^{(0)}}{\mu_k^{(0)}} \\
& = \tilde{J}_\infty^D(0|\x_0) + \tilde{J}_\infty^R(0|\x_0) = 1,
\end{align*}
where we used Eq. (\ref{eq:jinf_R}).  The last equality follows from
that the fact that $\tilde{J}_\infty^D(0|\x_0)$ and
$\tilde{J}_\infty^R(0|\x_0)$ are the splitting probabilities (e.g.,
$\tilde{J}_\infty^D(0|\x_0)$ is the probability of hitting $\pa_D$
before hitting $\pa_R$) so that their sum is equal to $1$.

The integral of $J_D(\ell,t|\x_0)$ over $\ell$ determines the
(marginal) probability density function of the escape time $T_0$
\begin{equation}
J_0^D(t|\x_0) = \int\limits_0^\infty d\ell \, J_D(\ell,t|\x_0).
\end{equation}
Integrating the continuity relation (\ref{eq:rho_cont}) over $\ell$
from $0$ to infinity and using the identity (\ref{eq:jR_int}), one
sees that
\begin{equation}  \label{eq:dS0_J0D}
\partial_t S_0(t|\x_0) = - J_0^D(t|\x_0),
\end{equation}
where we also used the normalization (\ref{eq:rho_norm}).  We thus
re-derived that the probability density of the first-passage time to
$\pa_D$ is obtained as the negative time derivative of the survival
probability $S_0(t|\x_0)$.  Expectedly, the derivation of this
classical quantity does not require the encounter-based approach.  

In turn, the integral of $J_D(\ell,t|\x_0)$ over $t$ yields the
(marginal) probability density of the boundary local time
$\ell_{T_0}^R$ acquired before the escape:
\begin{align}  
\rho_D(\ell|\x_0) & = \int\limits_0^\infty dt\, J_D(\ell,t|\x_0) = \tilde{J}_D(\ell,0|\x_0) \\  \nonumber
& = \tilde{J}_\infty^D(0|\x_0) \delta(\ell) + \sum\limits_{k=0}^\infty [V_k^{(0)}(\x_0)]^* \, e^{-\mu_k^{(0)}\ell} \, C_k^{(0)} .
\end{align}
Here, the first term accounts for the trajectories that never hit the
target region $\pa_R$ and moved directly to the escape region (with
the splitting probability $\tilde{J}_\infty^D(0|\x_0)$) so that the
acquired boundary local time is zero.  In turn, the second term
includes the contribution of remaining trajectories.  

The expression (\ref{eq:JD_tilde}) allows one to determine joint
positive integer-order moments of $\ell_{T_0}^R$ and $T_0$ as
\begin{equation}  \label{eq:ellT_T}
\E_{\x_0} \{ [\ell_{T_0}^R]^m  \, T_0^n \} = (-1)^n m!  \lim\limits_{p\to 0} \frac{\partial^n}{\partial p^n}
\sum\limits_{k=0}^\infty  \frac{[V_k^{(p)}(\x_0)]^* C_k^{(p)}}{[\mu_k^{(p)}]^{m+1}} 
\end{equation}
for any $m = 1,2,\ldots$ and any $n = 0,1,\ldots$ (for $m = 0$, the
contribution from the first (singular) term in Eq. (\ref{eq:JD_tilde})
has to be included).  In this way, one can evaluate not only the
moments of $\ell_{T_0}^R$ and $T_0$, but also correlations between
them.  In particular, the mean of $\ell_{T_0}^R$ reads
\begin{align}  \nonumber
\E_{\x_0} \{ \ell_{T_0}^R \} & = \sum\limits_{k=0}^\infty [V_k^{(0)}(\x_0)]^* \frac{C_k^{(0)}}{[\mu_k^{(0)}]^2} \\  \nonumber
& = \sum\limits_{k=0}^\infty [V_k^{(0)}(\x_0)]^* \frac{1}{\mu_k^{(0)}} \int\limits_{\pa_R} d\x \, v_k^{(0)}(\x) \\
& = \int\limits_{\pa_R} d\x \, D \tilde{G}_0(\x,0|\x_0),
\end{align}
where we used the identity (\ref{eq:G0_Ginf}).  This result is rather
intuitive.  In fact, the Green's function $D \tilde{G}_0(\x,0|\x_0)$
is known to be the mean residence time of Brownian motion in a point
$\x$ before the escape \cite{Morters}.  The integral of this quantity
over a thin layer of width $a$ near the target region $\pa_R$ yields
the mean residence time in this layer, while the rescaling by $D/a$
results in the mean boundary local time in the limit $a\to 0$, see
Eq. (\ref{eq:ellt_def}).

To evaluate higher-order moments, we first integrate
Eq. (\ref{eq:P_spectral}) $m+1$ times to get
\begin{align} \nonumber
& \sum\limits_{k=0}^\infty \frac{[V_k^{(p)}(\x_0)]^* \, V_k^{(p)}(\x)}{[\mu_k^{(p)}]^{m+1}} \\  \nonumber
& \quad = \int\limits_0^\infty d\ell_1 \int\limits_{\ell_1}^\infty d\ell_2 \, \ldots \int\limits_{\ell_{m+1}}^\infty d\ell_{m+1} 
D \tilde{P}(\x,\ell_{m+1},p|\x_0) \\
& \quad= \frac{1}{m!} \int\limits_0^\infty d\ell \, \ell^m \, D \tilde{P}(\x,\ell,p|\x_0)
\end{align}
for any $m = 1,2,\ldots$.  In turn, differentiating $m$ times the
fundamental relation (\ref{eq:Gq_P}) with respect to $q$, we derive
the following identity:
\begin{equation}
\sum\limits_{k=0}^\infty \frac{[V_k^{(p)}(\x_0)]^* \, V_k^{(p)}(\x)}{[\mu_k^{(p)}]^{m+1}}
 = \frac{(-1)^m}{m!} \lim\limits_{q\to 0} \partial_q^m D \tilde{G}_q(\x,p|\x_0) 
\end{equation}
for $m = 1,2,\ldots$ (for $m = 0$, there is an additional term, see
Eq. (\ref{eq:G0_Ginf})).  This identity provides a peculiar
interpretation of the derivatives of the Green's function
$\tilde{G}_q(\x,p|\x_0)$ with respect to $q$.  Using
Eqs. (\ref{eq:Ck_def}, \ref{eq:ellT_T}), one gets
\begin{align}  \label{eq:ellT_T2}
\E_{\x_0} \{ [\ell_{T_0}^R]^m \, T_0^n\} & = \frac{(-1)^{m+n+1}}{m!} \\   \nonumber
& \times \lim\limits_{q\to 0 \atop p\to 0} \partial^{m}_q 
\partial^n_p \int\limits_{\pa_D} d\x \,D \partial_n \tilde{G}_q(\x,p|\x_0).
\end{align}
The definitions (\ref{eq:jq_def}, \ref{eq:JqD_def}) further simplify
this relation as
\begin{equation}  \label{eq:ellT_T3}
\E_{\x_0} \{ [\ell_{T_0}^R]^m \, T_0^n\} = (-1)^{m+n}
\lim\limits_{q\to 0 \atop p\to 0} \partial^{m}_q \partial^n_p \tilde{J}_q^D(p|\x_0).
\end{equation}
Note that this relation is also applicable for $m = 0$, in which case
it reduces to the standard expression for the moments of the
first-passage time $T_0$:
\begin{equation}  \label{eq:Tn}
\E_{\x_0} \{ T_0^n\} = (-1)^{n} \lim\limits_{p\to 0} \partial^n_p \tilde{J}_0^D(p|\x_0).
\end{equation}
Like $\tilde{J}_0^D(p|\x_0)$ was known to be the generating function
of the first-passage time $T_0$, $\tilde{J}_q^D(p|\x_0)$ turns out to
be the generating function of both $T_0$ and $\ell_{T_0}^R$.  This is
not surprising given that $p$ and $q$ appear as the conjugate
variables in the Laplace transforms with respect to $t$ and $\ell$.
The elegant relation (\ref{eq:ellT_T3}) highlights the similarity
between the physical time $t$, which can be seen as a proxy for the
number of ``elementary jumps'' of the particle in the bulk, and the
boundary local time $\ell$, which is a proxy of the number of its
``jumps'' on the target region $\pa_R$.  This similarity was already
discussed in \cite{Grebenkov20} and particularly in
\cite{Grebenkov20b}, in which a surface-hopping propagator, based on
the boundary local time, was introduced.

\subsection{Reaction events}
\label{sec:jR}

Now we can inspect the restriction of the probability flux density
$j(\x,\ell,t|\x_0)$ to the target region $\pa_R$.  In the Laplace
domain, the spectral expansion (\ref{eq:P_spectral}) implies
\begin{align}  \label{eq:jRtilde}
\tilde{j}(\x,\ell,p|\x_0) &= \tilde{j}_\infty(\x,p|\x_0) \delta(\ell) \\  \nonumber
&- \sum\limits_{k=0}^\infty [V_k^{(p)}(\x_0)]^* \, \mu_k^{(p)}  v_k^{(p)}(\x) e^{-\mu_k^{(p)}\ell}  
\end{align}
for any $\x\in\pa_R$.  While this relation allows one to compute this
quantity, its probabilistic interpretation remains tricky, in
particular, due to its negative values according to
Eq. (\ref{eq:jR_int}).

To clarify this point, we first rewrite Eq. (\ref{eq:Gq_R}) as
$DG_q(\x,t|\x_0) = j_q(\x,t|\x_0)/q$ on $\pa_R$ and take its inverse
Laplace transform with respect to $q$ to get
\begin{equation}  \label{eq:P_intjR}
D P(\x,\ell,t|\x_0) = \int\limits_0^{\ell} d\ell'  \, j(\x,\ell',t|\x_0)    \quad (\x\in\pa_R).
\end{equation}
The probability flux density $j(\x,\ell',t|\x_0)$ on $\pa_R$ can thus
also be seen as the derivative of the full propagator on $\pa_R$ with
respect to $\ell$.  This is consistent with the boundary value problem
for the full propagator discussed in \cite{Grebenkov20}.  In the limit
$\ell\to\infty$, Eq. (\ref{eq:P_intjR}) reduces to the relation
(\ref{eq:jR_int}) because $P(\x,\ell,t|\x_0) \to 0$, i.e., the
probability of getting infinitely large values of the boundary local
time $\ell_t^R$ at a finite time $t$ is zero.  As the left-hand side
of Eq. (\ref{eq:P_intjR}) is a probability density, the integral in
the right-hand side is nonnegative, despite eventual negative values
of $j(\x,\ell',t|\x_0)$.  What does it represent?

In order to get its probabilistic interpretation, we introduce the
first-crossing time $\T_\ell$ of a threshold $\ell$ by the boundary
local time $\ell_t^R$:
\begin{equation}
\T_\ell = \inf\{ t>0 ~:~ \ell_t^R > \ell\} .
\end{equation}
If the particle has escaped the domain before crossing the threshold,
the first-crossing time is set to infinity.  We have then
\begin{equation}  \label{eq:Prob_Tell}
\P\{ \T_\ell > t\} = \P\{ \ell_t^R < \ell , ~ T_0 > t\} + \P\{ \ell_{T_0}^R < \ell, ~T_0 < t\} .
\end{equation}
The first term describes no crossing of the threshold $\ell$ when the
escape occurs after time $t$.  In turn, the second term describes the
escape event before $t$, for which the acquired boundary local time
$\ell_{T_0}^R$ remains below the threshold.  The probability density
of the first-crossing time reads then as
\begin{align*}
U(\ell,t|\x_0) & = - \partial_t \P\{\T_\ell > t\} \\
& = - \partial_t \int\limits_0^{\ell} d\ell' \, \rho(\ell',t|\x_0)
- \int\limits_0^{\ell} d\ell' \, J_D(\ell',t|\x_0) ,
\end{align*}
where the probability density $\rho(\ell',t|\x_0)$ was used to
evaluate the first term in Eq. (\ref{eq:Prob_Tell}), and the joint
probability density $J_D(\ell,t|\x_0)$ of $\ell_{T_0}^R$ and $T_0$ for
the second term.  Note that the definition of $\rho(\ell',t|\x_0)$
automatically accounts for no escape up to time $t$.  Using the
continuity equation (\ref{eq:rho_cont}), we finally get
\begin{equation}  \label{eq:U_JR}
U(\ell,t|\x_0) = \int\limits_0^{\ell} d\ell'  \, J_R(\ell',t|\x_0).
\end{equation}
Despite the fact that $J_R(\ell',t|\x_0)$ cannot be interpreted as a
joint probability density (in analogy with $J_D(\ell,t|\x_0)$), its
integral over $\ell'$ yields the probability density of the
first-crossing time $\T_\ell$.

According to Eq. (\ref{eq:P_intjR}), one gets another representation
\begin{equation}  \label{eq:U_Pint}
U(\ell,t|\x_0) = \int\limits_{\pa_R} d\x \, D P(\x,\ell,t|\x_0),
\end{equation}
which was earlier derived in \cite{Grebenkov20} for the particular
case $\pa_R = \pa$ (i.e., without the escape region).  This relation
has an intuitive interpretation.  Let us again consider a thin layer
of width $a$ near $\pa_R$.  By definition, the integral of the full
propagator $P(\x,\ell,t|\x_0)$ over $\x\in \pa_R$, multiplied by $a$
and $d\ell$, is the probability of finding the particle in that layer
at time $t$ with the boundary local time $\ell_t^R$ belonging to
$(\ell,\ell+d\ell)$.  As $\ell_t^R$ is an nondecreasing process that
increments only when $\X_t\in \pa_R$, the value $\ell$ is thus
achieved for the first time at $t$, i.e.,
\begin{align}  \nonumber
a\, d\ell \int\limits_{\pa_R} d\x \, P(\x,\ell,t|\x_0) & = \P_{\x_0}\{ \T_\ell \in (t,t+dt)\} \\
& = U(t,\ell|\x_0) dt .
\end{align}
Since the increments $d\ell$ and $dt$ are related as $d\ell = D dt/a$
according to Eq. (\ref{eq:elltR_def}), one gets Eq. (\ref{eq:U_Pint}).

In the Laplace domain, Eq. (\ref{eq:jRtilde}) implies
\begin{align*}
\tilde{U}(\ell,p|\x_0) 
& = \sum\limits_{k=0}^\infty [V_k^{(p)}(\x_0)]^* \,  e^{-\mu_k^{(p)}\ell} \int\limits_{\pa_R} d\x \, v_k^{(p)}(\x)  .
\end{align*}
Setting $\ell = 0$ and employing Eq. (\ref{eq:jinf_R}), one gets 
\begin{align*}
\tilde{U}(0,p|\x_0) & = \sum\limits_{k=0}^\infty [V_k^{(p)}(\x_0)]^* \, \int\limits_{\pa_R} d\x \, v_k^{(p)}(\x) \\
& = \int\limits_{\pa_R} d\x \, \tilde{j}_\infty(\x,p|\x_0) = \tilde{J}_\infty^R(p|\x_0),
\end{align*}
i.e.,
\begin{equation}  \label{eq:U0}
U(0,t|\x_0) = J_\infty^R(t|\x_0).
\end{equation}
In other words, as the first crossing of the threshold $\ell = 0$
corresponds to the first arrival onto $\pa_R$, the first-crossing time
$\T_0$ is simply the first-passage time to $\pa_R$ (before escaping
the domain).

As earlier stressed in \cite{Grebenkov20}, the probability density
$U(\ell,t|\x_0)$ of the first-crossing time is tightly related to the
probability density $J_q^R(t|\x_0)$ of the reaction time $\tau_q$ on
$\pa_R$ discussed in Sec. \ref{sec:conventional}.  In fact, using
Eqs. (\ref{eq:JqR_def}, \ref{eq:JR}, \ref{eq:jq_j}), we first get
\begin{equation}
J_q^R(t|\x_0) = \int\limits_0^\infty d\ell \, e^{-q\ell} \, J_R(\ell,t|\x_0) .
\end{equation}
In order to transform $J_R(\ell,t|\x_0)$ into $U(\ell,t|\x_0)$, one
needs to integrate by parts.  However, $J_R(\ell,t|\x_0)$ contains a
singular term $J_\infty^R(t|\x_0) \delta(\ell)$ that has to be treated
separately.  We have then
\begin{align*}
J_q^R(t|\x_0) & = J_\infty^R(t|\x_0) + \int\limits_{0^+}^\infty d\ell \, e^{-q\ell} \, J_R(\ell,t|\x_0) \\
& = J_\infty^R(t|\x_0) + \underbrace{\left.\biggl(\int\limits_{0^+}^\infty d\ell \, e^{-q\ell} \, 
U(\ell,t|\x_0)\biggr)\right|_{\ell=0}^\infty}_{-U(0,t|\x_0)} \\
& + q \int\limits_{0^+}^\infty d\ell \, e^{-q\ell} \, U(\ell,t|\x_0) ,
\end{align*}
where we wrote the lower bound of the integral as $0^+$ to highlight
that the singular term was excluded.  According to Eq. (\ref{eq:U0}),
two first terms cancel each other, yielding
\begin{equation}  \label{eq:JqR_U}
J_q^R(t|\x_0) = \int\limits_{0}^\infty d\ell \, q\, e^{-q\ell} \, U(\ell,t|\x_0) .
\end{equation}

To interpret this relation, we introduce a random threshold
$\hat{\ell}$ obeying the exponential probability law with the rate
$q$, $\P\{ \hat{\ell} > \ell\} = e^{-q\ell}$, so that $q e^{-q\ell}$
is the probability density of this law.  As a consequence, the
integral (\ref{eq:JqR_U}) is the average over random realizations of
the threshold $\hat{\ell}$ of the probability density of the
first-crossing time $\T_{\hat{\ell}}$.  Since the left-hand side is
the probability density of the reaction time $\tau_q$, we conclude
that
\begin{equation}  \label{eq:tau_def}
\tau_q = \T_{\hat{\ell}} = \inf\{ t > 0 ~:~ \ell_t^R > \hat{\ell} \} ,
\end{equation}
i.e., the reaction occurs when the boundary local time $\ell_t^R$
exceeds the random threshold $\hat{\ell}$ with the exponential law.
We emphasize that the exponential law follows directly from the
postulated Robin boundary condition (\ref{eq:Gq_R}).  This
interpretation, earlier suggested in \cite{Grebenkov20}, allows one to
go beyond the Robin boundary condition and to implement various
surface reaction mechanisms by choosing an appropriate law for the
random threshold $\hat{\ell}$.  In this way, our definition
(\ref{eq:tau_def}) of the reaction time $\tau_q$ remains valid in a
much more general setting, while Eq. (\ref{eq:JqR_U}), which was
specific to the Robin boundary condition, is then generalized to
\begin{equation}  \label{eq:JqR_U_psi}
J_\psi^R(t|\x_0) = \int\limits_{0}^\infty d\ell \, \psi(\ell) \, U(\ell,t|\x_0) ,
\end{equation}
where $\psi(\ell)$ is the chosen probability density of the random
threshold $\hat{\ell}$.  Different choices of $\psi(\ell)$ and its
consequences on the distribution of the reaction time were discussed
in \cite{Grebenkov20}.  This extension of surface reaction mechanisms
is directly applicable to our setting with the escape region $\pa_D$.
In this way, we made a step further towards more realistic modeling of
diffusion-controlled reactions by incorporating the effect of escape
events.

We complete this section by providing a deeper probabilistic
interpretation of the first-passage time $T_q$ to the escape region
$\pa_D$ in the presence of the reactive region $\pa_R$.  As discussed
in Sec. \ref{sec:conventional}, this random variable is described by
the probability density $J_q^D(t|\x_0)$, which is related to
$J_D(\ell,t|\x_0)$ due to Eq. (\ref{eq:jq_j}) as
\begin{equation}  \label{eq:JqD_JD}
J_q^D(t|\x_0) = \int\limits_0^\infty d\ell \, e^{-q\ell} \, J_D(\ell,t|\x_0).
\end{equation}
As previously, the factor $e^{-q\ell}$ incorporates the condition of
no reaction until the escape, which was automatically included into
the left-hand side via the Robin boundary condition on $\pa_R$.  In
turn, if the reaction happens before the escape, the escape time $T_q$
is set to infinity.  In other words, one has
\begin{equation}
T_q = \left\{ \begin{array}{ll} T_0 & \textrm{if}~ \ell_{T_0}^R < \hat{\ell} , 
\\  +\infty & \textrm{otherwise.} \end{array} \right.
\end{equation}
As earlier, the dependence on the reactivity parameter $q$ is
incorporated via the random threshold $\hat{\ell}$ obeying the
exponential law with the rate $q$.  This representation highlights the
effect of surface reactions onto the fisrt-passage time $T_q$.  In
fact, if the target region $\pa_R$ was inert, the particle would reach
the escape region $\pa_D$ at a random time $T_0$.  In turn, the
reactivity of $\pa_R$ makes the only change that the particle can
react on $\pa_R$ and thus never escape.  If the boundary local time
$\ell_{T_0}^R$ at $T_0$ has not crossed the random threshold
$\hat{\ell}$, the reaction on $\pa_R$ has not happened, and $T_q =
T_0$.  In contrast, if the threshold $\hat{\ell}$ has been crossed,
the reaction occurred before $T_0$ so that $T_q = \infty$.  The above
definition of the first-passage time to the escape region $\pa_D$
within the encounter-based approach allows one to study this quantity
for other surface reaction mechanisms, beyond the conventional one
described by Robin boundary condition.  For this purpose, one can
choose an appropriate law $\P\{\hat{\ell} > \ell\}$ for the random
threshold $\hat{\ell}$ and replace the factor $e^{-q\ell}$ in
Eq. (\ref{eq:JqD_JD}) by this law.  Further investigations of this
setting present an interesting perspective of the present work.

\section{Spherical target}
\label{sec:sphere}

In order to illustrate the general properties of diffusion-controlled
reactions with eventual escape, we consider diffusion between two
concentric spheres of radii $R$ and $L$: $\Omega = \{ \x\in\R^3 ~:~ R
< |\x| < L\}$.  The inner sphere represents the target region $\pa_R$,
while the outer sphere is the escape region $\pa_D$ (in this setting,
$\pa_N = \emptyset$).  In spherical coordinates $\x =
(r,\theta,\phi)$, the modified Helmholtz equation can be solved via
separation of variables.  In particular, the eigenbasis of the
Dirichlet-to-Neumann operator is well known
\cite{Grebenkov20b,Grebenkov20c,Grebenkov22c}.  In fact, the
rotational invariance of the domain implies that the eigenfunctions of
$\M_p$ are the (normalized) spherical harmonics $Y_{mn}(\theta,\phi)$
\begin{equation}
v_{nm} = \frac{1}{R} Y_{mn}(\theta,\phi)  \qquad \left({n = 0,1,\ldots, \atop |m| \leq n} \right),
\end{equation}
where the double index $nm$ is used instead of a single index $k$.  In
turn, the eigenvalues are
\begin{equation}
\mu_{nm}^{(p)} = - \bigl(\partial_r g_n^{(p)}(r)\bigr)_{r=R} \,,
\end{equation}
where 
\begin{equation}
g_n^{(p)}(r) = \frac{k_n(\alpha L) i_n(\alpha r) - i_n(\alpha L) k_n(\alpha r)}
{k_n(\alpha L) i_n(\alpha R) - i_n(\alpha L) k_n(\alpha R)}
\end{equation}
are the radial functions, with $\alpha = \sqrt{p/D}$ and $i_n(z)$ and
$k_n(z)$ being the modified spherical Bessel functions of the first
and second kind.  Note that $g_n^{(p)}(L) = 0$ and $g_n^{(p)}(R) = 1$.
One sees that the eigenvalues are $(2n+1)$ times degenerate (they do
not depend on the index $m$), while the eigenfunctions do not depend
on $p$.  Finally, it is easy to check that
\begin{equation}
V_{nm}^{(p)}(\x) = g_n^{(p)}(r) \, v_{nm}(\theta,\phi).
\end{equation}
As most quantities of interest are obtained by integrating over the
target region $\pa_R$, the orthogonality of eigenfunctions $v_{nm}$ to
$v_{00} = 1/\sqrt{4\pi R^2}$ ensures that all terms vanish except this
ground eigenmode.  Since $i_0(z) = \sinh(z)/z$ and $k_0(z) =
e^{-z}/z$, one finds
\begin{equation}
g_0^{(p)}(r) = \frac{R \, \sinh(\alpha(L-r))}{r \, \sinh(\alpha(L-R))}
\end{equation}
and
\begin{equation}
\mu_0^{(p)} = \mu_{00}^{(p)} = \frac{1}{R} + \alpha \, \ctanh(\alpha(L-R)) .
\end{equation}
Note that
\begin{align*}
& \int\limits_{\Omega} d\x \, V_{00}^{(p)}(\x) = \frac{4\pi}{\sqrt{4\pi R^2}} \int\limits_R^L dr \, r^2 \, g_0^{(p)}(r) \\
& = \frac{\sqrt{4\pi}}{\sinh(\alpha(L-R))} \biggl(-\frac{L}{\alpha} \\
& + \frac{\alpha R \cosh(\alpha(L-R)) + \sinh(\alpha(L-R))}{\alpha^2} \biggr)
\end{align*}
and
\begin{equation*}
C_0^{(p)} = \sqrt{4\pi}\, \frac{\alpha L}{\sinh(\alpha(L-R))} \,.
\end{equation*}

One can also compute
\begin{widetext}
\begin{subequations}
\begin{align}
J_q^D(p|\x_0) & = \frac{L}{r_0} \, \frac{(1 + qR) \sinh(\alpha(r_0-R)) + \alpha R \cosh(\alpha(r_0-R))}
{(1 + qR) \sinh(\alpha(L-R)) + \alpha R \cosh(\alpha(L-R))} \,, \\
J_q^R(p|\x_0) & = \frac{R}{r_0} \, \frac{qR \sinh(\alpha(L-r_0))}
{(1 +qR)\sinh(\alpha(L-R)) + \alpha R \cosh(\alpha(L-R)) } \,, 
\end{align}
\end{subequations}
\end{widetext}
where $r_0 = |\x_0|$, from which 
\begin{subequations}
\begin{align*}
J_\infty^R(p|\x_0) & = \frac{R \, \sinh(\alpha(L-r_0))}{r_0 \, \sinh(\alpha(L-R))}  \,,\\
J_\infty^D(p|\x_0) & = \frac{L \sinh(\alpha(r_0-R))}{r_0 \sinh(\alpha(L-R))} \,, \\
J_0^D(p|\x_0) & = \frac{L}{r_0} \, \frac{\sinh(\alpha(r_0-R)) + \alpha R \cosh(\alpha(r_0-R))}
{\sinh(\alpha(L-R)) + \alpha R \cosh(\alpha(L-R))} \,.
\end{align*}
\end{subequations}

In the following illustrations, we fix units of length and time by
setting $R = 1$ and $D = 1$.

\subsection{Escape events}

According to Eq. (\ref{eq:JD_tilde}), we find 
\begin{align}   \nonumber
\tilde{J}_D(\ell,p|\x_0) 
& = \frac{L \sinh(\alpha(r_0-R))}{r_0 \sinh(\alpha(L-R))} \delta(\ell)  \\  \label{eq:JD_sphere}
&+ \frac{L \, \alpha \, \sinh(\alpha(L-r_0))}{r_0 \, \sinh^2(\alpha(L-R))} \, e^{-\mu_0^{(p)}\ell} \,,
\end{align}
from which an inverse Laplace transform with respect to $p$ yields the
joint probability density $J_D(\ell,t|\x_0)$ of $\ell_{T_0}^R$ and
$T_0$.  The inversion of the first term in Eq. (\ref{eq:JD_sphere})
can be found explicitly via the residue theorem.  In turn, the
presence of the $p$-dependent factor $e^{-\mu_0^{(p)}\ell}$ in the
second term makes the inversion challenging (see \cite{Grebenkov20c}
for some analytical tools).  For this reason, we employed a numerical
Laplace transform inversion by the Talbot algorithm \cite{Talbot79}.

\begin{figure}[t!]
\begin{center}
\includegraphics[width=88mm]{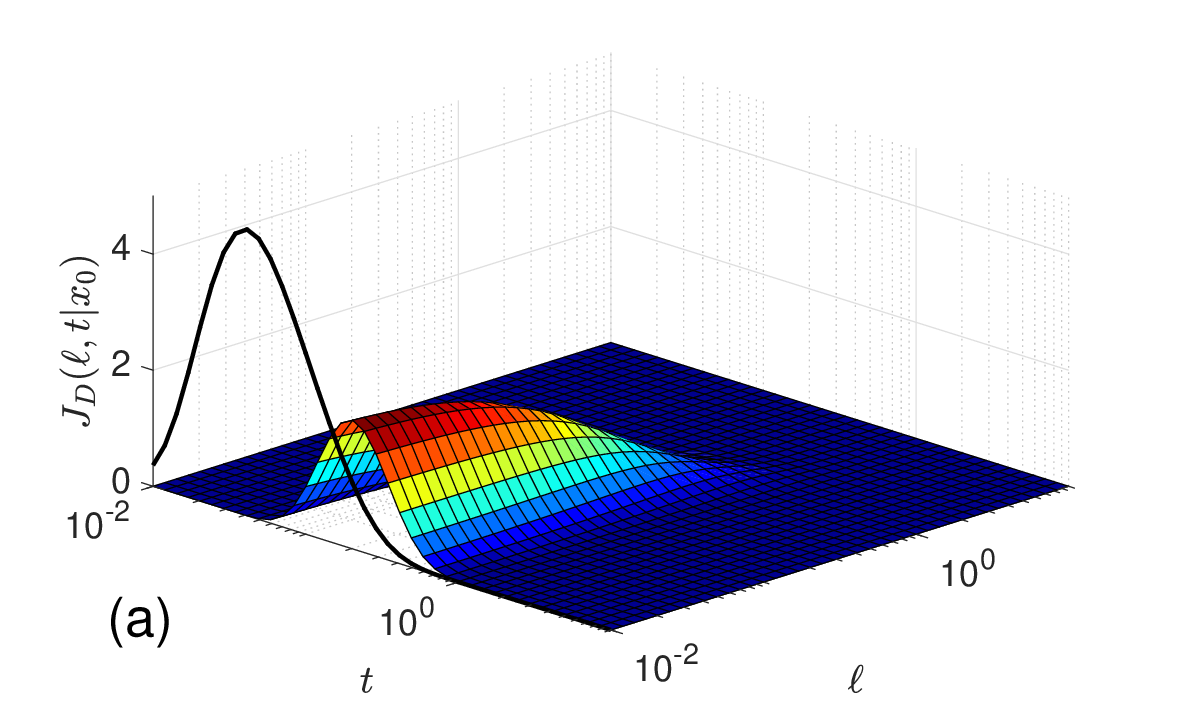} 
\includegraphics[width=88mm]{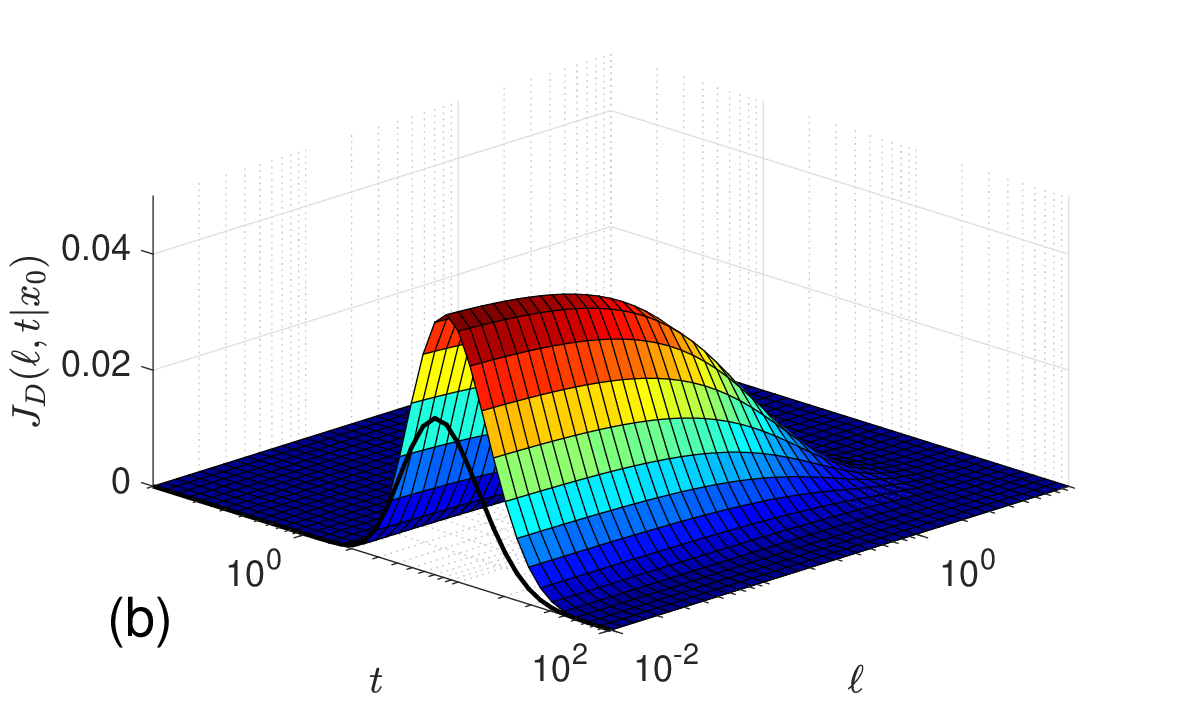} 
\end{center}
\caption{
Joint probability density $J_D(\ell,t|\x_0)$ of the boundary local
time $\ell_{T_0}^R$ and the escape time $T_0$ for diffusion between
two concentric spheres of radii $R$ and $L$, with $R = 1$, $r_0 =
1.5$, $D = 1$, $L = 2$ {\bf (a)} and $L = 10$ {\bf (b)}.  Surface
shows the regular part of this density, while black solid line
presents the prefactor $J_\infty^D(t|\x_0)$ in front of the singular
term $\delta(\ell)$.  Note that this curve should be located at $\ell
= 0$, which is not visible on the logarithmic scale; it was thus
artificially put at $\ell = 10^{-2}$ for illustration purposes.  The
ranges of times $t$ differ by factor $10$ between two panels.}
\label{fig:JD}
\end{figure}

Figure \ref{fig:JD} illustrates the behavior of $J_D(\ell,t|\x_0)$.
Two terms in Eq. (\ref{eq:JD_sphere}) give respectively singular and
regular contributions to this density with respect to $\ell$.  The
regular contribution as a function of $\ell$ and $t$ is shown by a
surface.  In turn, the singular contribution corresponding to $\ell =
0$ is shown by a black line.  For the panel {\bf (a)}, we set $L = 2$
and $r_0 = 1.5$, i.e., the particle starts in the middle between the
target region at $R = 1$ and the escape region at $L = 2$; in turn, in
the panel {\bf (b)}, the escape region is moved to $L = 10$.

Since the particle needs to diffuse from its starting point $\x_0$ to
the escape region at $L$, its escape at very short times is highly
unlikely.  In particular, the probability density $J_\infty^D(t|\x_0)$
of the first-passage time to $\pa_D$ determining the singular term, is
known to exhibit the short-time behavior of a L\'evy-Smirnov form
$t^{-3/2} e^{-(L-r_0)^2/(4Dt)}$ (see \cite{Godec16,Grebenkov18} and
references therein).  Similar behavior can be derived for the regular
contribution.  In the same vein, the probability of escape at very
long times is also negligible because the particle cannot avoid the
escape region too long.  These arguments explain a distinct maximum of
$J_D(\ell,t|\x_0)$ with respect to $t$ at intermediate times.
Finally, too large values of the boundary local time $\ell_{T_0}^R$
for intermediate escape times would require for the particle to stay
too long near the target region $\pa_R$, which is also highly
improbable.  One sees therefore that $J_\infty^D(t|\x_0)$ exhibits a
single ``boss'' that rapidly goes down as $t\to 0$, $t\to
\infty$, or $\ell\to \infty$.

The panel {\bf (a)} of Fig. \ref{fig:JD} indicates that the maximum of
the singular contribution (the curve $J_\infty^D(t|\x_0)$) is shifted
to shorter times with respect to the boss of $J_D(\ell,t|\x_0)$.  In
fact, the particle that never hit the target region $\pa_R$ (and thus
has $\ell_{T_0}^R = 0$) needs to travel the distance $L-r_0$ from the
starting point to the escape region that determines the maximum of
$J_\infty^D(t|\x_0)$.  In turn, the particle that first encountered
$\pa_R$ and then escaped, has to travel the longer distance $(r_0 - R)
+ (L-R)$ that requires longer time.  When the starting point gets
closer to $\pa_D$, the separation between two maxima is even larger,
while the location of $\x_0$ near $\pa_R$ would reduce this
separation.  Similarly, if $L$ is much larger than $r_0$ (as on the
panel {\bf (b)}), the difference between two distances is relatively
small, and two maxima are close.

\begin{figure}[t!]
\begin{center}
\includegraphics[width=88mm]{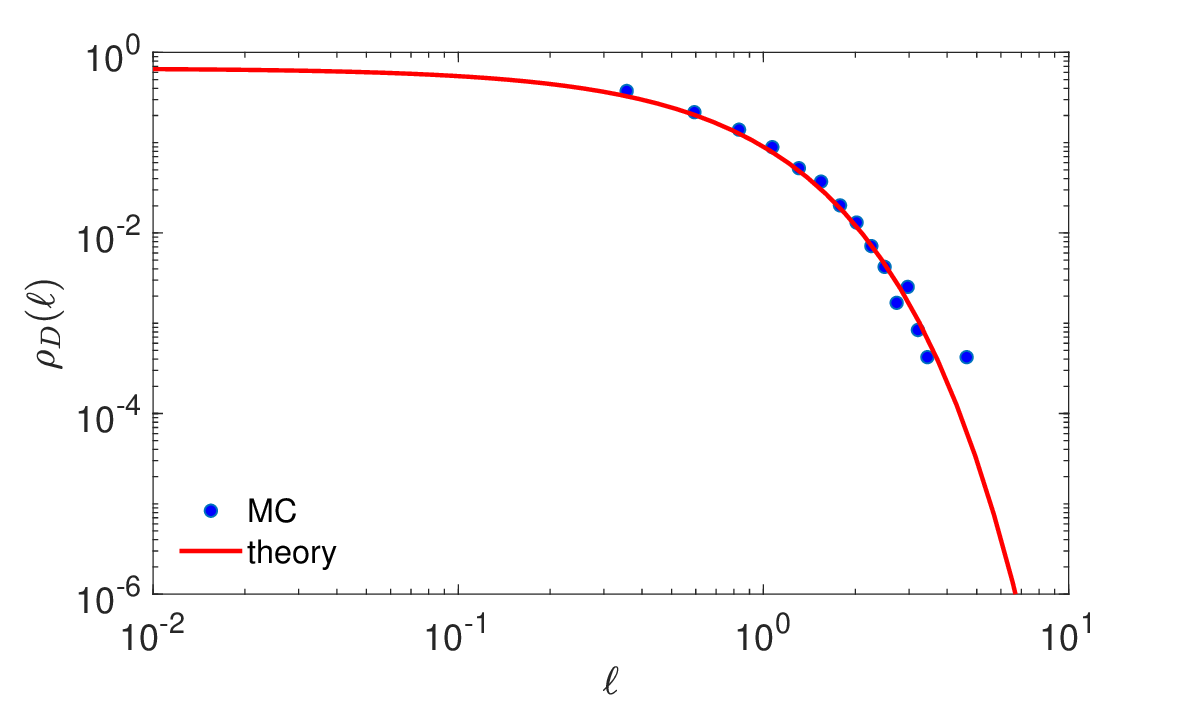} 
\end{center}
\caption{
The regular part of the probability density $\rho_D(\ell|\x_0)$ of the
acquired boundary local time $\ell_{T_0}^R$ up to the escape time
$T_0$ for diffusion between two concentric spheres of radii $R$ and
$L$, with $R = 1$, $L = 2$, $r_0 = 1.5$, and $D = 1$.  Solid line
shows the exact solution (\ref{eq:rhoD_sphere}), while symbols present
a renormalized histogram from Monte Carlo simulations with $M = 10^4$
particles, the time step $\delta = 10^{-4}$ and the layer width $a =
5\sqrt{2D\delta}$.}
\label{fig:rhoD}
\end{figure}

Setting $p = 0$ in Eq. (\ref{eq:JD_sphere}), one gets the (marginal)
probability density of $\ell_{T_0}^R$:
\begin{align}  \nonumber
\rho_D(\ell|\x_0) & = \frac{L}{r_0(L-R)} \biggl((r_0-R) \delta(\ell) \\  \label{eq:rhoD_sphere}
& +  \frac{L-r_0}{L-R} \, e^{-\ell(1/R + 1/(L-R))} \biggr).
\end{align}
This is a mixture of an exponential distribution and an atom at $\ell
= 0$.  
Figure \ref{fig:rhoD} illustrates the regular (exponential) part of
this distribution for the example from Fig. \ref{fig:JD}(a).  As an
additional validation step, we also realized Monte Carlo simulations
of random trajectories and thus independently evaluated the statistics
of $\ell_{T_0}^R$.  For this purpose, a random trajectory of a
particle was simulated by adding independent Gaussian increments,
with mean zero and variance $\sqrt{2D\delta}$ in each direction, where
$\delta$ is a small time step.  If the particle jumps inside the inner
sphere of radius $R$, it is normally reflected back.  At each step
when the particle is within a layer of width $a$ near the inner sphere
(i.e., when $|\X_t|<R+a$), the boundary local time $\ell_t^R$ is
incremented by $D\delta/a$ according to Eq. (\ref{eq:elltR_def}).  The
simulation is stopped when the particle crosses the outer sphere of
radius $L$.  Repeating such a simulation $M$ times, one gets an
empirical statistics of the escape time $T_0$ and of the acquires
boundary local time $\ell_{T_0}^R$.

Figure \ref{fig:rhoD} confirms an excellent agreement between the
exact solution (\ref{eq:rhoD_sphere}) and simulations.  One can also
appreciate that getting the behavior of $\rho_D(\ell|\x_0)$ at too
small or too large $\ell$ is problematic for Monte Carlo simulations.
In fact, estimations at small $\ell$ require very short time steps in
the modeling of the random trajectory (and thus too long simulations).
In turn, estimations at large $\ell$ require too many simulated
trajectories, as the probability of getting large $\ell_{T_0}^R$ is
exponentially small.

The moments of $\ell_{T_0}^R$ are particularly simple:
\begin{equation}
\E_{\x_0} \{ [\ell_{T_0}^R]^m\} = m! \, \frac{R(L-r_0)}{r_0(L-R)} \biggl(\frac{1}{R} + \frac{1}{L-R}\biggr)^{-m} .
\end{equation}
In particular,
\begin{subequations}
\begin{align}
\E_{\x_0} \{ \ell_{T_0}^R \} & = R^2(1/r_0 - 1/L)  \,, \\
\V_{\x_0}\{ \ell_{T_0}^R \} & = R^3(1/r_0 - 1/L)(2 - R/r_0 - R/L),
\end{align}
\end{subequations}
where $\V_{\x_0}$ denotes the variance.  The moments of $T_0$ are
determined via Eq. (\ref{eq:Tn}); in particular, one has
\begin{equation}
\E_{\x_0}\{ T_0 \} = \frac{(L-r_0)\bigl( r_0 L(r_0+L) - 2R^3 \bigr)}{6Dr_0 L} 
\end{equation}
and
\begin{align}  \nonumber
& \V_{\x_0} \{ T_0 \} = \frac{L-r_0}{90 D^2 L^2 r_0^2} \biggl(r_0^2 L^5 + L^4 r_0^3 + r_0^4 L^3 + r_0^5 L^2 \\
& \quad - 20L^2 r_0^2 R^3  + 36R^5 r_0 L -10R^6L - 10R^6r_0\biggr).
\end{align}
In turn, the joint moments can be found from
\begin{align}  \nonumber
& \E_{\x_0} \{ [\ell_{T_0}^R]^m \, T_0^n \}  = (-1)^n m! \frac{L}{r_0} \\  
& \quad \times \lim\limits_{p\to 0} \frac{\partial^n}{\partial p^n}
\biggl(\frac{\alpha \, \sinh(\alpha(L-r_0))}{\sinh^2(\alpha(L-R))} [\mu_0^{(p)}]^{-m-1}\biggr) .
\end{align}
For instance, we get
\begin{align}  \nonumber
\E_{\x_0} \{ \ell_{T_0}^R \, T_0 \} & = \frac{R^2(L-r_0)}{6D L^2 r_0} \biggl(2(L+2R)(L-R)^2 \\
& - L(L-r_0)^2\biggr)  \,.
\end{align}
Expectedly, all these quantities vanish as $r_0 \to L$ because the
particle started on $\pa_D$ escapes immediately, yielding $T_0 =
\ell_{T_0}^R = 0$ in a deterministic way.  The above expressions allow
one to compute the Peason's correlation coefficient between $T_0$ and
$\ell_{T_0}^R$:
\begin{equation}
C = \frac{\E_{\x_0} \{ \ell_{T_0}^R \, T_0 \} - \E_{\x_0} \{ \ell_{T_0}^R\} \, \E_{\x_0} \{ T_0 \}}
{\sqrt{\V_{\x_0} \{ T_0 \}}  \, \sqrt{\V_{\x_0} \{ \ell_{T_0}^R \} }} \,.
\end{equation}

\begin{figure}[t!]
\begin{center}
\includegraphics[width=88mm]{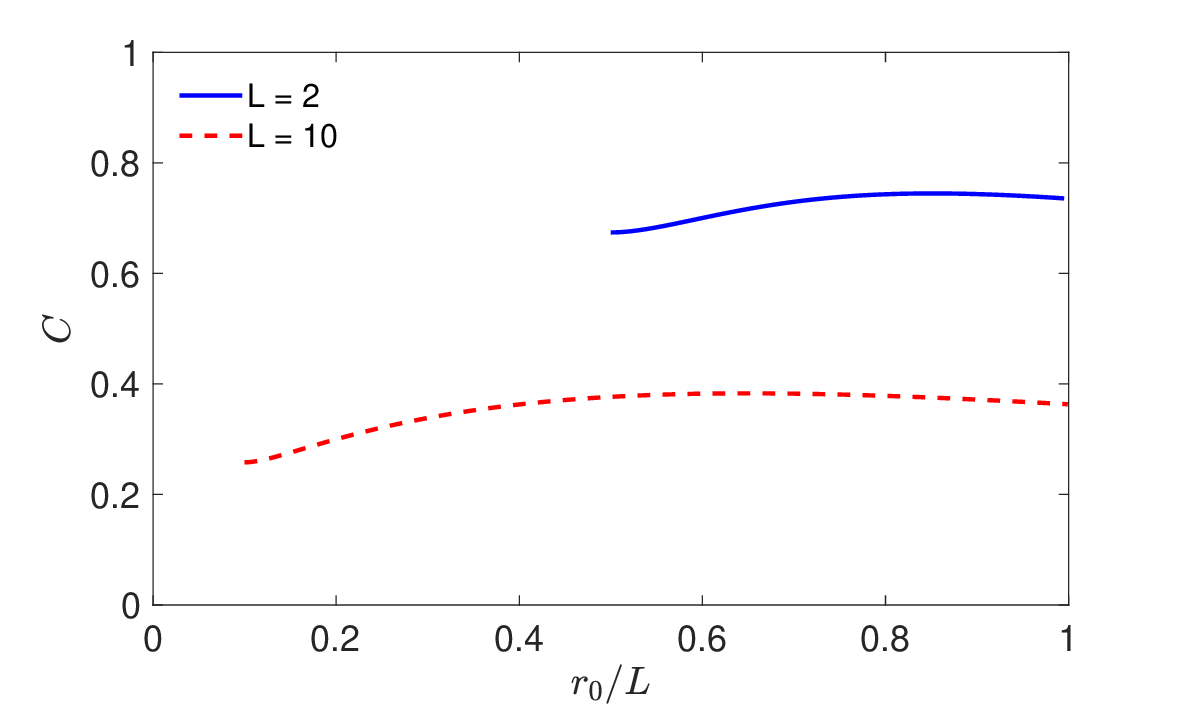} 
\end{center}
\caption{
The correlation coefficient between the escape time $T_0$ and the
boundary local time $\ell_{T_0}^R$ as a function of $r_0/L$, for
diffusion between two concentric spheres of radii $R$ and $L$, with $R
= 1$, $D = 1$, and two values of $L$ indicated in the legend.  }
\label{fig:C}
\end{figure}

Figure \ref{fig:C} shows the correlation coefficient $C$ as a function
of $r_0/L$ for two values of $L$.  In both cases, the correlation is
positive.  Indeed, if the particle escapes faster (at smaller $T_0$),
it would generally have lower chances to encounter the target region
frequently so that the boundary local time $\ell_{T_0}^R$ would also
be smaller.  Note that correlations are higher for the case $L = 2$
than for $L = 10$.  In other words, when the distance between the
escape and target regions is larger, the particle would generally take
longer time to diffuse in the confining domain before the escape and
thus to decorrelate these random variables.  Even though all moments
vanish as $r_0 \to L$, the correlation coefficient gets a nontrivial
limit.  Curiously, both curves exhibit slightly non-monotonous
behavior with respect to $r_0$.

\subsection{First-crossing times}

According to Eq. (\ref{eq:jRtilde}), we have
\begin{equation}   \label{eq:JR_sphere}
\tilde{J}_R(\ell,p|\x_0) = g_0(r_0) \biggl(\delta(\ell) - \mu_0^{(p)} e^{-\mu_0^{(p)}\ell}\biggr)
\end{equation}
and thus
\begin{equation}  \label{eq:Up_sphere}
\tilde{U}(\ell,p|\x_0) = g_0(r_0) \, e^{-\mu_0^{(p)} \ell} .
\end{equation}

\begin{figure}[t!]
\begin{center}
\includegraphics[width=88mm]{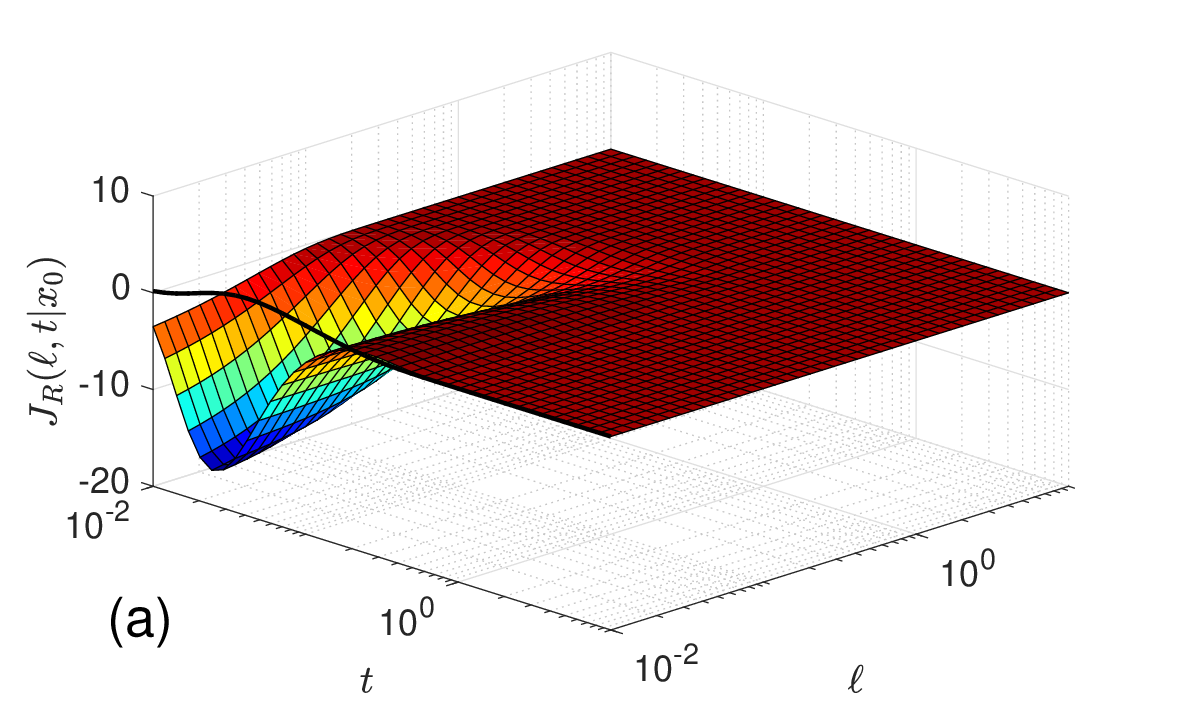} 
\includegraphics[width=88mm]{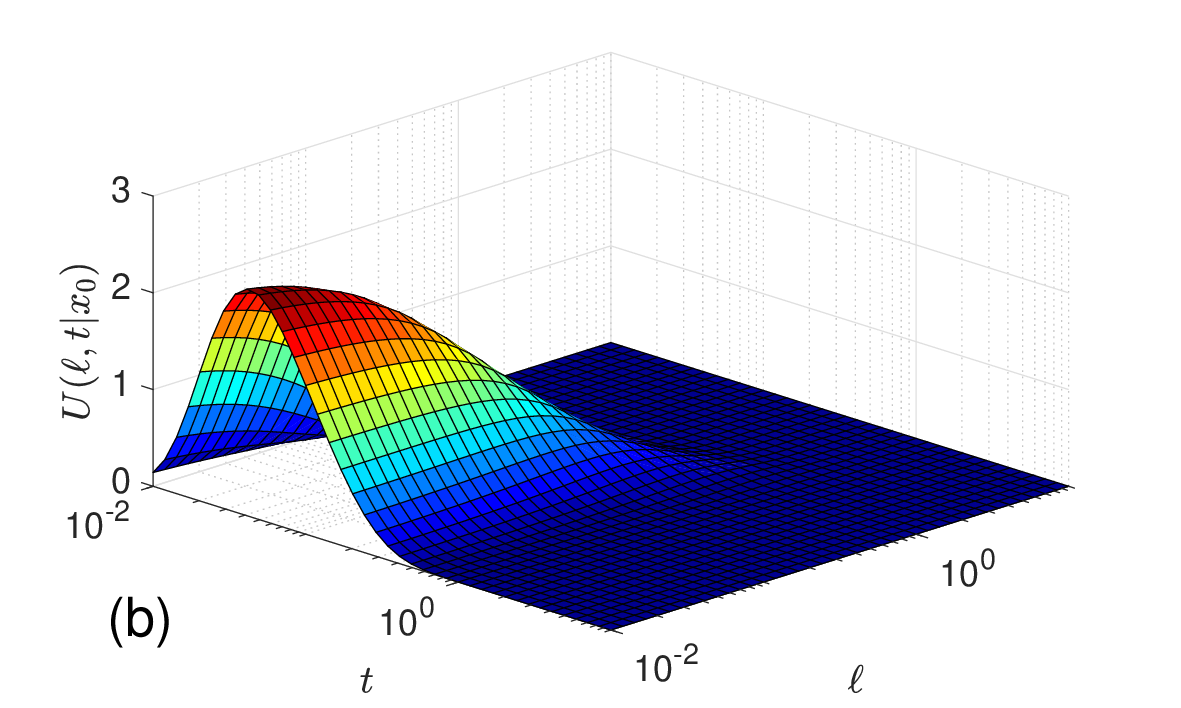} 
\end{center}
\caption{
{\bf (a)} Probability flux density $J_R(\ell,t|\x_0)$ for diffusion
between two concentric spheres of radii $R$ and $L$, with $R = 1$, $L
= 2$, $r_0 = 1.5$, and $D = 1$.  Surface shows the regular part of
this density, which was obtained by the numerical Laplace transform
inversion of Eq. (\ref{eq:JR_sphere}) with respect to $p$ by Talbot
algorithm.  Black solid line presents the prefactor
$J_\infty^R(t|\x_0)$ in front of the singular term $\delta(\ell)$.
Note that this curve should be located at $\ell = 0$, which is not
visible on the logarithmic scale; it was thus artificially put at
$\ell = 10^{-2}$ for illustration purposes.  {\bf (b)} Probability
density $U(\ell,t|\x_0)$ of the first-crossing time $\T_\ell$ for the
same setting.  This density is obtained by the numerical Laplace
transform inversion of Eq. (\ref{eq:Up_sphere}) with respect to $p$ by
Talbot algorithm.  According to Eq. (\ref{eq:U_JR}),
$J_R(\ell,t|\x_0)$ is the derivative of $U(\ell,t|\x_0)$ with respect
to $\ell$.}
\label{fig:JR}
\end{figure}

Figure \ref{fig:JR}(a) illustrates the behavior of $J_R(\ell,t|\x_0)$.
As previously, we plot the regular and singular parts by a surface and
a black curve.  In sharp contrast to $J_D(\ell,t|\x_0)$ shown on
Fig. \ref{fig:JD}, the regular part of $J_R(\ell,t|\x_0)$ is negative,
as discussed in Sec. \ref{sec:jR}.  Moreover, the minimum of the
regular part is not shifted with respect to the maximum of the
singular term $J_\infty^R(t|\x_0)$.  This is related to the fact that
both extrema are determined by the time needed for the particle to
travel from $\x_0$ to the target region at $R$.  Interestingly, the
``amplitude'' of the regular part of $J_R(\ell,t|\x_0)$ is an order of
magnitude higher than that of $J_D(\ell,t|\x_0)$.  In particular, the
black curve showing $J_\infty^R(t|\x_0)$ looks almost flat at this
scale.  This observation does not contradict Eq. (\ref{eq:U_JR}) that
ensures the positivity of the integral of $J_R(\ell',t|\x_0)$ over
$\ell'$ from $0$ to any $\ell$.  This is confirmed by
Fig. \ref{fig:JR}(b) showing this integral.  For each value of the
threshold $\ell$, this figure gives the probability density
$U(\ell,t|\x_0)$ of the first-crossing time $\T_\ell$.  We recall that
this density is not normalized to $1$ due to escape events:
\begin{align}   \label{eq:U_sphere_norm}
\int\limits_0^\infty dt \, U(\ell,t|\x_0) & = \tilde{U}(\ell,0|\x_0) \\  \nonumber
& = \frac{R(L-r_0)}{r_0(L-R)} e^{-\ell/(1/R+1/(L-R))} \leq 1.
\end{align}
The probability density $U(\ell,t|\x_0)$ as a function of both $\ell$
and $t$ exhibits a single ``boss''; in fact, it is unlikely to cross a
given threshold $\ell$ at too short or too long times; in turn, the
decrease at large $\ell$ is ensured by the normalization relation
(\ref{eq:U_sphere_norm}).  Even though the shape of $U(\ell,t|\x_0)$
looks similar to that of $J_D(\ell,t|\x_0)$ shown on
Fig. \ref{fig:JD}, their probabilistic interpretations are different.

Yet another difference between $J_D(\ell,t|\x_0)$ and
$J_R(\ell,t|\x_0)$ is that the former strongly depends on the location
of the escape region (compare two panels of Fig. \ref{fig:JD}),
whereas the latter exhibits only a weak dependence on $L$ whenever $L$
is large enough.  For this reason, we do not present the graphs of
$J_R(\ell,t|\x_0)$ and $U(\ell,t|\x_0)$ for $L = 10$, because they are
almost indistinguishable from that shown on Fig. \ref{fig:JR} for $L =
2$.  This can be seen from Eq. (\ref{eq:Up_sphere}): when $\sqrt{p/D}
(L-r_0) \gg 1$, one has
\begin{equation}  \label{eq:Up_approx}
\tilde{U}(\ell,p|\x_0) \approx \frac{R}{r_0} e^{-(r_0-R + \ell)\sqrt{p/D} - \ell/R}  \,,
\end{equation}
which is independent of $L$.  Its inverse Laplace transform yields
\begin{equation}   \label{eq:Ut_approx}
U(\ell,t|\x_0) \approx \frac{R e^{-\ell/R}}{r_0}\, \frac{(r_0-R + \ell) e^{-(r_0-R + \ell)^2/(4Dt)}}{\sqrt{4\pi Dt^3}}  \,.
\end{equation}
The right-hand side is actually the exact form of the probability
density $U(\ell,t|\x_0)$ for a spherical target in the
three-dimensional space (i.e., in the limit $L\to\infty$), as reported
in \cite{Grebenkov20b}.  As the approximation (\ref{eq:Up_approx})
fails in the limit $p\to 0$ (in which case $\sqrt{p/D} (L-r_0) \gg 1$
cannot hold), the approximation (\ref{eq:Ut_approx}) fails at long
times.  This is expected because if the particle has enough time to
explore the domain $\Omega$, it will unavoidably hit the escape
region.

\begin{figure}[t!]
\begin{center}
\includegraphics[width=88mm]{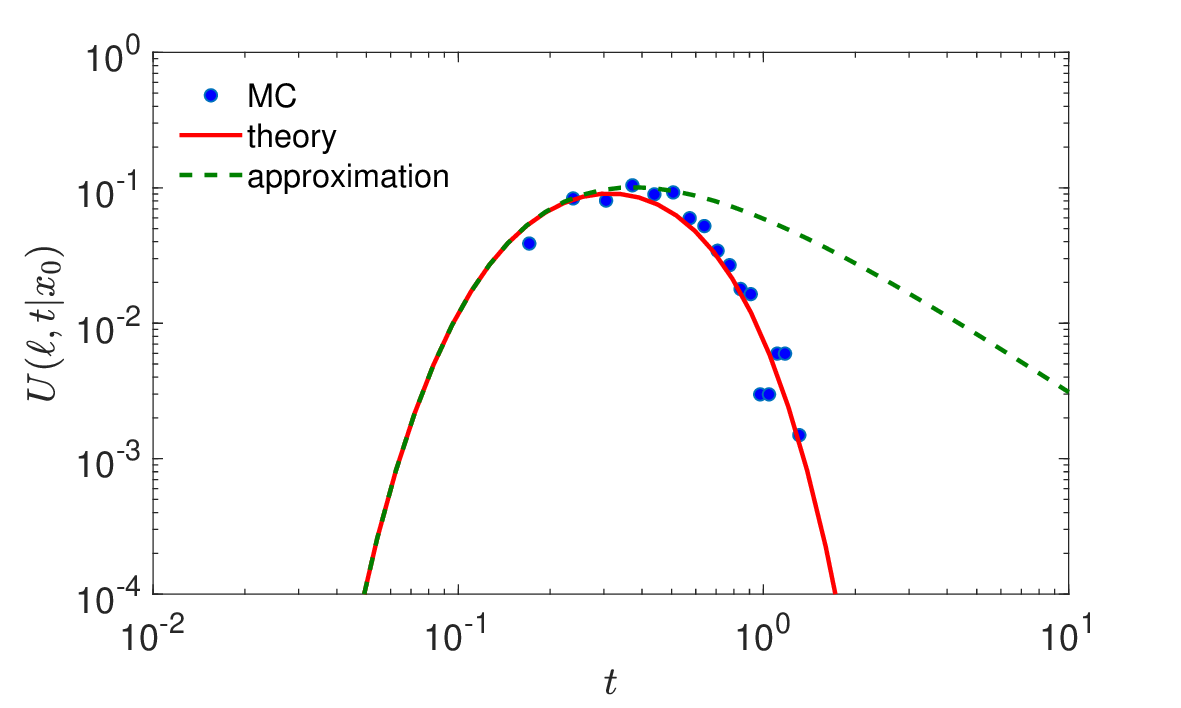} 
\end{center}
\caption{
Probability density $U(\ell,t|\x_0)$ of the first-crossing time
$\T_\ell$ of a threshold $\ell$ by the boundary local time $\ell_t^R$
for diffusion between two concentric spheres of radii $R$ and $L$,
with $R = 1$, $L = 2$, $r_0 = 1.5$, $D = 1$, and $\ell = 1$.  Symbols
present a renormalized histogram from Monte Carlo simulations with $M
= 10^4$, $\delta = 10^{-4}$ and $a = 5\sqrt{2Dt\delta}$, solid line
shows the numerical Laplace transform inversion of
Eq. (\ref{eq:Up_sphere}) by Talbot algorithm, while dashed line
indicates the approximation (\ref{eq:Ut_approx}) corresponding to the
limit $L \to \infty$.}
\label{fig:U}
\end{figure}

Figure \ref{fig:U} allows one to compare the probability density
$U(\ell,t|\x_0)$ for a particular value $\ell = 1$ with the results of
Monte Carlo simulations.  One observes an excellent agreement between
the theory and simulations.  This figure also highlights the
difficulty in getting the values of $U(\ell,t|\x_0)$ at short and long
times by Monte Carlo simulations.  We also plot the approximation
(\ref{eq:Ut_approx}) in the limit $L \to \infty$.  As said earlier,
this approximation is very accurate at short times but fails at long
times.  As $L$ increases, the validity range of the approximation
progressively extends to longer and longer times.

\section{Conclusion}
\label{sec:discussion}

In this paper, we revised the encounter-based approach to
diffusion-mediated surface phenomena and generalized this formalism by
allowing a generic partition of the boundary $\pa$ into three parts: a
target region $\pa_R$, a reflecting region $\pa_N$ and an escape
region $\pa_D$.  While the original formulation in \cite{Grebenkov20}
dealt with the whole boundary as the target ($\pa = \pa_R$), our
generalization brings a greater flexibility to modeling
diffusion-controlled reactions and covers a broad scope of related
first-passage problems.  From the mathematical point of view, this
extension essentially consists in adding Neumann and Dirichlet
boundary conditions on $\pa_N$ and $\pa_D$, respectively.  Despite
this apparent simplicity, the inclusion of a ``killing mechanism'' for
the diffusing particle (an escape through $\pa_D$) has required some
conceptual refinements and appropriate modifications in the formalism.
In particular, we focused in the paper on the probability flux density
$j(\x,\ell,t|\x_0)$, which was mostly ignored in former works.

On the one hand, the restriction of $j(\x,\ell,t|\x_0)$ to the escape
region $\pa_D$ determined, for the first time, the joint probability
density of the escape time $T_0$, the particle position $\X_{T_0}$,
and the acquired boundary local time $\ell_{T_0}^R$ on the target
region $\pa_R$.  In turn, the marginal density $J_D(\ell,t|\x_0)$, in
which the position $\X_{T_0}$ was averaged out, allows one to
characterize not only the statistics of the number of encounters
between the target region and the diffusing particle before its
escape, but also correlations with the escape time.  In particular, we
obtained three equivalent representations (\ref{eq:ellT_T},
\ref{eq:ellT_T2}, \ref{eq:ellT_T3}) of the joint moments of
$\ell_{T_0}^R$ and $T_0$ and revealed a probabilistic interpretation
of multiple derivatives of the Green's function
$\tilde{G}_q(\x,p|\x_0)$ with respect to $p$ and $q$.  Note that
earlier works dealt exclusively with the escape time itself and were
mainly focused on the mean value and its dependence on the geometric
and kinetic parameters.  However, it is important to emphasize that
the mean escape time, which is mostly affected by long but rare
trajectories, may be orders of magnitude larger than the typical time
of the escape process, and thus be noninformative or even misleading
\cite{Grebenkov18,Mattos12,Godec16b,Grebenkov18b,Reva21}.  For
instance, this may occur when the particle starts in a neighborhood of
a small escape region.  In this situation, the whole distribution of
the escape time (or another quantity such as $\ell_{T_0}^R$) is needed
to characterize the escape process.  For this reason, we focused on
the probability flux densities that were much less studied in the
past.

On the other hand, the restriction of $j(\x,\ell,t|\x_0)$ to the
target region $\pa_R$ required a subtle probabilistic interpretation
in terms of the first-crossing time $\T_\ell$ of a given threshold
$\ell$ by the boundary local time $\ell_t^R$.  In particular, the
integral of $j(\x,\ell',t|\x_0)$ over $\x\in\pa_R$ and over $\ell'$
from $0$ to $\ell$ determined the probability density $U(\ell,t|\x_0)$
of $\T_\ell$.  The latter played a key role in the generalization of
conventional surface reactions, described by the Robin boundary
condition, to more general mechanisms.  While this generalization was
already presented in \cite{Grebenkov20}, its extension in the presence
of escape events required some refinements.  For instance, the former
derivation relied on the nondecreasing character of the boundary local
time $\ell_t$ and its immediate consequence that $\P\{ \T_\ell > t\} =
\P\{ \ell_t < \ell\}$.  In the presence of the escape region, this
relation has to be replaced by Eq. (\ref{eq:Prob_Tell}) accounting for
the value $\T_\ell = \infty$ in the case if the particle has escaped
before crossing the threshold $\ell$.  More generally, the lack of a
proper normalization of probability densities had to be carefully
managed.

For illustrative purposes, we considered diffusion between concentric
spheres, the inner sphere of radius $R$ being the target region and
the outer sphere of radius $L$ being the escape region.  This basic
example allowed us to compute all the discussed quantities in a simple
compact form in the Laplace domain and then to evaluate the inverse
Laplace transform numerically.  We presented both $J_D(\ell,t|\x_0)$
and $J_R(\ell,t|\x_0)$, as well as some derived quantities such as
$\rho_D(\ell|\x_0)$ and $U(\ell,t|\x_0)$.  We discussed two cases: $L
= 2$ (when the target and escape regions are relatively close to each
other), and $L = 10$ (when they are well separated).  Expectedly, the
joint probability density $J_D(\ell,t|\x_0)$ strongly depends on the
location of the escape region, whereas $J_R(\ell,t|\x_0)$ showed only
a weak dependence on $L$.  In particular, the probability density
$U(\ell,t|\x_0)$ approaches the well-known explicit form
(\ref{eq:Ut_approx}) for a spherical target in the three-dimensional
space.

While we considered ordinary diffusion inside the confining domain,
the presented formalism allows one to easily incorporate a first-order
reaction kinetics in the bulk.  This kinetics may account for a
spontaneous disappearance of the diffusing particle (or its
``activity'') with a constant rate $\gamma$ due to radioactive
disintegration, photobleaching, relaxation of its excited state,
disassembly, failure, or biological death.  Whatever the actual
killing mechanism, such a ``mortal'' particle can be treated as having
a random lifetime $\delta$ that obeys the exponential law $\P\{ \delta
> t \} = e^{-\gamma t}$
\cite{Yuste13,Meerson15,Grebenkov17f,Meerson19}.  In the conventional
approach, the killing mechanism in the bulk is included by adding the
term $-\gamma G_q(\x,t|\x_0)$ to the right-hand side of the diffusion
equation (\ref{eq:Gq_diff}).  Its Laplace transform with respect to
$t$ yields the same modified Helmholtz equation (\ref{eq:Gq_Helm}), in
which $p$ is replaced by $p' = p +\gamma$.  As a consequence, most
results that we obtained in the Laplace domain, remain valid up to
this trivial change.  Moreover, many Laplace-transformed quantities,
evaluated at $p = \gamma$, admit useful probabilistic interpretations
in terms of the stopping condition at the ``death'' time $\delta$.
For instance,
\begin{equation}
\gamma\, \tilde{\rho}(\ell,\gamma|\x_0) = \int\limits_0^\infty dt \, \gamma e^{-\gamma t} \, \rho(\ell,t|\x_0)
\end{equation}
is the probability density of the boundary local time $\ell_\delta^R$
at the random time $\delta$ of the particle ``death'' in the bulk,
with $\gamma e^{-\gamma t}$ being the probability density of $\delta$.
In other words, many Laplace-transformed quantities that we derived in
this paper, have their own interest, even without evaluating their
inverse Laplace transforms.  The combined effect of the first-order
kinetics in the bulk, surface reactions on the target region, and
escape events can be further explored.

The present work can be extended in several directions.  On the
mathematical side, the spectral properties of the Dirichlet-to-Neumann
operator need further attention.  In addition to a rigorous
demonstration of the announced basic properties, the asymptotic
behavior of the eigenvalues $\mu_k^{(p)}$ has to be uncovered, in
particular, in the limit when either $\pa_R$ or $\pa_D$ (or both) is
small.  We expect that some matched asymptotic tools
\cite{Bressloff22a,Bressloff22b} can be adapted to investigate this
problem.  On the application side, one can investigate in more detail
the effect of escape events onto various surface reaction mechanisms
introduced in \cite{Grebenkov20}.  The proposed extension can be
further analyzed in the presence of multiple independently diffusing
particles \cite{Grebenkov22b}, eventual resetting mechanisms
\cite{Evans20,Bressloff22c,Benkhadaj22} and drifts
\cite{Grebenkov22a}.  Future applications of the extended
encounter-based approach can bring more realistic features to a
theoretical description of biology-inspired transport processes,
notably in living cells.

\begin{acknowledgments}
The author thanks the Alexander von Humboldt Foundation for support
through the Bessel Prize award.
\end{acknowledgments}


\begin{thebibliography}{10}


\bibitem{Alberts}		B. Alberts, A. Johnson, J. Lewis, D. Morgan, M. Raff, K. Roberts, and P. Walter, 
				{\it Molecular Biology of the Cell} 
				(Garland Science, New York, NY, 2014).

\bibitem{Rice}			S. A. Rice, 
				{\it Diffusion-limited reactions} 
				(Elsevier, Amsterdam, 1985).

\bibitem{Lauffenburger}		D. A. Lauffenburger and J. Linderman, 
				{\it Receptors: Models for Binding, Trafficking, and Signaling}
				(Oxford University Press, Oxford, 1993).

\bibitem{Schuss}		Z. Schuss, 
				{\it Brownian Dynamics at Boundaries and Interfaces in Physics, Chemistry and Biology}
				(Springer, New York, 2013).

\bibitem{Lindenberg}		K. Lindenberg, R. Metzler, and G. Oshanin 
				{\it Chemical Kinetics: Beyond the Textbook}
				(World Scientific, New Jersey, 2019).


\bibitem{Bressloff13} 		P. C. Bressloff and J. M. Newby, 
				Stochastic models of intracellular transport,
				Rev. Mod. Phys. {\bf 85}, 135-196 (2013).





\bibitem{Redner} 		S. Redner, 
				{\it A Guide to First Passage Processes}
				(Cambridge, Cambridge University press, 2001).

\bibitem{Metzler} 		R. Metzler, G. Oshanin, and S. Redner (Eds), 
				{\it First-Passage Phenomena and Their Applications}
				(Singapore, World Scientific, 2014).


\bibitem{Condamin07}		S. Condamin, O. B\'enichou, V. Tejedor, R. Voituriez, and J. Klafter,
				First-passage time in complex scale-invariant media,
				Nature {\bf 450}, 77 (2007).

\bibitem{Benichou10}		O. B\'enichou, C. Chevalier, J. Klafter, B. Meyer, and R. Voituriez, 
				Geometry-controlled kinetics,
				Nat. Chem. {\bf 2}, 472-477 (2010).

\bibitem{Holcman13} 		D. Holcman and Z. Schuss,
				Control of flux by narrow passages and hidden targets in cellular biology,
				Phys. Progr. Rep. {\bf 76}, 074601 (2013).

\bibitem{Benichou14}		O. B\'enichou and R. Voituriez,
				From first-passage times of random walks in confinement to geometry-controlled kinetics,
				Phys. Rep. {\bf 539}, 225-284 (2014).

\bibitem{Grebenkov16}		D. S. Grebenkov, 
				Universal formula for the mean first passage time in planar domains, 
				Phys. Rev. Lett. {\bf 117}, 260201 (2016).

\bibitem{Guerin16}		T. Gu\'erin, N. Levernier, O. B\'enichou, and R. Voituriez,
				Mean first-passage times of non-Markovian random walkers in confinement,
				Nature {\bf 534}, 356-359 (2016).

\bibitem{Lanoiselee18}		Y. Lanoisel\'ee, N. Moutal, and D. S. Grebenkov, 
				Diffusion-limited reactions in dynamic heterogeneous media, 
				Nature Commun. {\bf 9}, 4398 (2018).




\bibitem{Gardiner}		C. W. Gardiner,
				{\it Handbook of stochastic methods for physics, chemistry and the natural sciences}
				(Springer: Berlin, 1985).

\bibitem{VanKampen}		N. G. Van Kampen,
				{\it Stochastic Processes in Physics and Chemistry}
				(Elsevier, Amsterdam, 1992).



\bibitem{Collins49}		F. C. Collins and G. E. Kimball, 
				Diffusion-controlled reaction rates, 
				J. Colloid Sci. {\bf 4}, 425 (1949).

\bibitem{Berg77}		H. C. Berg and E. M. Purcell,
				Physics of chemoreception,
				Biophys. J. {\bf 20}, 193 (1977).

\bibitem{Sano79}		H. Sano and M. Tachiya, 
				Partially diffusion-controlled recombination, 
				J. Chem. Phys. {\bf 71}, 1276 (1979).

\bibitem{Brownstein79}		K. R. Brownstein and C. E. Tarr, 
				Importance of Classical Diffusion in NMR Studies of Water in Biological Cells, 
				Phys. Rev. A {\bf 19}, 2446-2453 (1979).

\bibitem{Weiss86}		G. H. Weiss, 
				Overview of theoretical models for reaction rates,
				J. Stat. Phys. {\bf 42}, 3 (1986).

\bibitem{Powles92}		J. G. Powles, M. J. D. Mallett, G. Rickayzen, and W. A. B. Evans, 
				Exact analytic solutions for diffusion impeded by an infinite array of partially 
				permeable barriers, 
				Proc. R. Soc. London A {\bf 436}, 391-403 (1992).

\bibitem{Sapoval94}		B. Sapoval, 
				General Formulation of Laplacian Transfer Across Irregular Surfaces, 
				Phys. Rev. Lett. {\bf 73}, 3314 (1994).

\bibitem{Sapoval02}		B. Sapoval, M. Filoche, and E. Weibel, 
				Smaller is better -- but not too small: A physical scale for the design 
				of the mammalian pulmonary acinus, 
				Proc. Nat. Ac. Sci. USA {\bf 99}, 10411-10416 (2002).

\bibitem{Grebenkov05}		D. S. Grebenkov, M. Filoche, B. Sapoval, and M. Felici, 
				Diffusion-reaction in Branched Structures: Theory and Application to the Lung Acinus, 
				Phys. Rev. Lett. {\bf 94}, 050602 (2005).

\bibitem{Traytak07}		S. D. Traytak and W. Price, 
				Exact solution for anisotropic diffusion-controlled reactions with partially reflecting conditions, 
				J. Chem. Phys. {\bf 127}, 184508 (2007).

\bibitem{Bressloff08}		P. C. Bressloff, B. A. Earnshaw, and M. J. Ward, 
				Diffusion of protein receptors on a cylindrical dendritic membrane with partially absorbing traps, 
				SIAM J. Appl. Math. {\bf 68}, 1223-1246 (2008).

\bibitem{Singer08}		A. Singer, Z. Schuss, A. Osipov, and D. Holcman, 
				Partially reflected diffusion,
				SIAM J. Appl. Math. {\bf 68}, 844 (2008).

\bibitem{Grebenkov10}		D. S. Grebenkov, 
				Searching for partially reactive sites: Analytical results for spherical targets, 
				J. Chem. Phys. {\bf 132}, 034104 (2010).

\bibitem{Lawley15}		S. D. Lawley and J. P. Keener,  
				A new derivation of Robin boundary conditions through homogenization of
				a stochastically switching boundary,
				SIAM J. Appl. Dyn. Syst. {\bf 14}, 1845-1867 (2015).

\bibitem{Grebenkov15}		D. S. Grebenkov, 
				Analytical representations of the spread harmonic measure density,
				Phys. Rev. E {\bf 91}, 052108 (2015).

\bibitem{Serov16}		A. S. Serov, C. Salafia, D. S. Grebenkov, and M. Filoche, 
				The Role of Morphology in Mathematical Models of Placental Gas Exchange, 
				J. Appl. Physiol. {\bf 120}, 17-28 (2016).

\bibitem{Bressloff17}		P. C. Bressloff, 
				Stochastic switching in biology: from genotype to phenotype,
				J. Phys. A. {\bf 50}, 133001 (2017).

\bibitem{Grebenkov17}		D. S. Grebenkov and G. Oshanin, 
				Diffusive escape through a narrow opening: new insights into a classic problem, 
				Phys. Chem. Chem. Phys. {\bf 19}, 2723-2739 (2017).

\bibitem{Piazza19}		F. Piazza and D. S. Grebenkov, 
				Diffusion-controlled reaction rate on non-spherical partially absorbing axisymmetric surfaces, 
				Phys. Chem. Chem. Phys. {\bf 21}, 25896 (2019).



\bibitem{Grebenkov19b}		D. S. Grebenkov, 
				Spectral theory of imperfect diffusion-controlled reactions on heterogeneous catalytic surfaces, 
				J. Chem. Phys. {\bf 151}, 104108 (2019).



\bibitem{Carslaw}		H. S. Carslaw and J. C. Jaeger,
				{\it Conduction of Heat in Solids}, 2nd Ed.
				(Oxford University Press, 1959).

\bibitem{Crank}			J. Crank, 
				{\it The Mathematics of Diffusion}
				(Oxford University Press, 1956).

\bibitem{Thambynayagam}		R. K. M. Thambynayagam,
				{\it The Diffusion Handbook: Applied Solutions for Engineers}
				(McGraw Hill, 2011).

\bibitem{Grebenkov13}		D. S. Grebenkov and B.-T. Nguyen, 
				Geometrical structure of Laplacian eigenfunctions, 
				SIAM Rev. {\bf 55}, 601-667 (2013).





\bibitem{Mazya85}		V. G. Maz'ya, S. A. Nazarov, and B. A. Plamenevskii,
				 Asymptotic expansions of the eigenvalues of boundary 
				value problems for the Laplace operator in domains with small holes,
				Math. USSR. Izv {\bf 24}, 321-345 (1985).

\bibitem{Ward93}		M. J. Ward and J. B. Keller, 
				Strong localized perturbations of eigenvalue problems,
				SIAM J. Appl. Math. {\bf 53}, 770-798 (1993).

\bibitem{Kolokolnikov05}	T. Kolokolnikov, M. S. Titcombe, and M. J. Ward,
				Optimizing the fundamental Neumann eigenvalue for the Laplacian in a domain with small traps,
				Eur. J. Appl. Math {\bf 16}, 161 (2005).

\bibitem{Singer06a}		A. Singer, Z. Schuss, D. Holcman, and R. S. Eisenberg,
				Narrow escape, part I,
				J. Stat. Phys. {\bf 122}, 437-463 (2006).

\bibitem{Singer06b}		A. Singer, Z. Schuss, and D. Holcman,
				Narrow escape, part II: the circular disk,
				J. Stat. Phys. {\bf 122}, 465 (2006).

\bibitem{Singer06c}		A. Singer, Z. Schuss, and D. Holcman,
				Narrow escape, part III: non-smooth domains and Riemann surfaces,
				J. Stat. Phys. {\bf 122}, 491 (2006).

\bibitem{Schuss07}		Z. Schuss, A. Singer, and D. Holcman,
				The narrow escape problem for diffusion in cellular microdomains,
				Proc. Nat. Acad. Sci. USA {\bf 104}, 16098-16103 (2007).

\bibitem{Benichou08}		O. B\'enichou and R. Voituriez,
				Narrow-escape time problem: time needed for a particle to exit a confining 
				domain through a small window,
				Phys. Rev. Lett. {\bf 100}, 168105 (2008).

\bibitem{Pillay10}		S. Pillay, M. J. Ward, A. Peirce, and T. Kolokolnikov, 
				An asymptotic analysis of the mean first passage time for narrow escape problems: 
				part I: two-dimensional domains,
				Multiscale Model. Simul. {\bf 8}, 803-835 (2010).

\bibitem{Cheviakov10}		A. F. Cheviakov, M. J. Ward, and R. Straube,
				An asymptotic analysis of the mean first passage time for narrow escape problems: 
				part II: the sphere,
				Multiscale Model. Simul. {\bf 8}, 836-870 (2010).

\bibitem{Cheviakov11}		A. F. Cheviakov and M. J. Ward,
				Optimizing the principal eigenvalue of the Laplacian in a sphere with interior traps,
				Math. Comput. Modelling {\bf 53}, 1394-1409 (2011).

\bibitem{Cheviakov12}		A. F. Cheviakov, A. S. Reimer, and M. J. Ward,
				Mathematical modeling and numerical computation of narrow escape problems,
				Phys. Rev. E {\bf 85}, 021131 (2012).

\bibitem{Holcman14}		D. Holcman and Z. Schuss, 
				The narrow escape problem,
				SIAM Rev. {\bf 56}, 213-257 (2014).

\bibitem{Agranov18}		T. Agranov and B. Meerson, 
				Narrow escape of interacting diffusing particles,
				Phys. Rev. Lett. {\bf 120}, 120601 (2018).

\bibitem{Grebenkov19c}		D. S. Grebenkov, R. Metzler, and G. Oshanin, 
				Full distribution of first exit times in the narrow escape problem, 
				New J. Phys. {\bf 21}, 122001 (2019).





\bibitem{Traytak92}		S. D. Traytak,
				The diffusive interaction in diffusion-limited reactions: the steady-state case,
				Chem. Phys. Lett. {\bf 197}, 247-254 (1992).

\bibitem{Condamin06}		S. Condamin and O. B\'enichou,
				Exact expressions of mean first-passage times and splitting probabilities 
				for random walks in bounded rectangular domains,
				J. Chem. Phys. {\bf 124}, 206103 (2006).

\bibitem{Chevalier11}		C. Chevalier, O. B\'enichou, B. Meyer and R. Voituriez,
				First-passage quantities of Brownian motion in a bounded domain with 
				multiple targets: a unified approach,
				J. Phys. A: Math. Theor. {\bf 44}, 025002 (2011).

\bibitem{Galanti16}		M. Galanti, D. Fanelli, S. D. Traytak, and F. Piazza,
				Theory of diffusion-influenced reactions in complex geometries
				Phys. Chem. Chem. Phys. {\bf 18}, 15950-15954 (2016).

\bibitem{Grebenkov19f}		D. S. Grebenkov and S. Traytak, 
				Semi-analytical computation of Laplacian Green functions in three-dimensional 
				domains with disconnected spherical boundaries, 
				J. Comput. Phys. {\bf 379}, 91-117 (2019).

\bibitem{Grebenkov20f}		D. S. Grebenkov, 
				Diffusion toward non-overlapping partially reactive spherical traps: 
				fresh insights onto classic problems, 
				J. Chem. Phys. {\bf 152}, 244108 (2020).

\bibitem{Klinger22}		J. Klinger, R. Voituriez, and O. B\'enichou,
				Splitting Probabilities of Symmetric Jump Processes,
				Phys. Rev. Lett. {\bf 129}, 140603 (2022).





\bibitem{Grebenkov20}		D. S. Grebenkov, 
				Paradigm Shift in Diffusion-Mediated Surface Phenomena, 
				Phys. Rev. Lett. {\bf 125}, 078102 (2020).




\bibitem{Levy}			P. L\'evy, 
				{\it Processus Stochastiques et Mouvement Brownien} 
				(Paris: Gauthier-Villard, 1965).

\bibitem{Ito}			K. It\^o and H. P. McKean, 
				{\it Diffusion Processes and Their Sample Paths}
				(Berlin: Springer, 1965).

\bibitem{Freidlin}		M. Freidlin,
				{\it Functional Integration and Partial Differential Equations}
				(Annals of Mathematics Studies, Princeton, NJ: Princeton University Press, 1985).



\bibitem{Borodin}		A. N. Borodin and P. Salminen,
				{\it Handbook of Brownian Motion: Facts and Formulae}
				(Birkhauser Verlag, Basel-Boston-Berlin, 1996).

\bibitem{Majumdar05}		S. N. Majumdar,
				Brownian functionals in physics and computer science,
				Curr. Sci. {\bf 88}, 2076-2092 (2005).


\bibitem{Grebenkov21}		D. S. Grebenkov, 
				Statistics of boundary encounters by a particle diffusing outside a compact planar domain, 
				J. Phys. A.: Math. Theor. {\bf 54}, 015003 (2021). 




\bibitem{Grebenkov20c}		D. S. Grebenkov, 
				Joint distribution of multiple boundary local times and related first-passage 
				time problems with multiple targets, 
				J. Stat. Mech. 103205 (2020).







\bibitem{Arendt14}		W. Arendt, A. F. M. ter Elst, J. B. Kennedy, and M. Sauter,
				The Dirichlet-to-Neumann operator via hidden compactness,
				J. Funct. Anal. {\bf 266}, 1757-1786 (2014).

\bibitem{Daners14}		D. Daners, 
				Non-positivity of the semigroup generated by the Dirichlet-to-Neumann operator,
				Positivity {\bf 18}, 235-256 (2014).

\bibitem{terElst14}		A. F. M. ter Elst and E. M. Ouhabaz,
				Analysis of the heat kernel of the Dirichlet-to-Neumann operator,
				J. Funct. Anal. {\bf 267}, 4066-4109 (2014).

\bibitem{Behrndt15}		J. Behrndt and A. F. M. ter Elst,
				Dirichlet-to-Neumann maps on bounded Lipschitz domains,
				J. Differ. Equ. {\bf 259}, 5903-5926 (2015).

\bibitem{Arendt15}		W. Arendt and A. F. M. ter Elst,
				The Dirichlet-to-Neumann operator on exterior domains,
				Potential Anal. {\bf 43}, 313-340 (2015).

\bibitem{Hassell17}		A. Hassell and V. Ivrii, 
				Spectral asymptotics for the semiclassical Dirichlet to Neumann operator
				J. Spectr. Theory {\bf 7}, 881-905 (2017).

\bibitem{Girouard17}		A. Girouard and I. Polterovich, 
				Spectral geometry of the Steklov problem,
				J. Spectr. Theory {\bf 7}, 321-359 (2017).








\bibitem{Grebenkov19a}		D. S. Grebenkov, 
				Probability distribution of the boundary local time of reflected Brownian 
				motion in Euclidean domains, 
				Phys. Rev. E {\bf 100}, 062110 (2019).




\bibitem{Morters}		P. M\"orters and Y. Peres,
				{\it Brownian Motion}
				(Cambridge Series in Statistical and Probabilistic Mathematics, 
				Cambridge University Press, New York, 2010).



\bibitem{Grebenkov20b}		D. S. Grebenkov, 
				Surface Hopping Propagator: An Alternative Approach to Diffusion-Influenced Reactions, 
				Phys. Rev. E {\bf 102}, 032125 (2020).

\bibitem{Grebenkov22c}		D. S. Grebenkov, 
				Statistics of diffusive encounters with a small target: Three complementary approaches, 
				J. Stat. Mech. 083205 (2022). 



\bibitem{Talbot79}		A. Talbot, 
				The accurate numerical inversion of Laplace transforms,
				J. Inst. Math. Appl. {\bf 23}, 97-120 (1979).


\bibitem{Godec16}		A. Godec and R. Metzler,
				Universal Proximity Effect in Target Search Kinetics in the Few-Encounter Limit,
				Phys. Rev. X {\bf 6}, 041037 (2016).

\bibitem{Grebenkov18}		D. S. Grebenkov, R. Metzler, and G. Oshanin, 
				Strong defocusing of molecular reaction times results from an 
				interplay of geometry and reaction control, 
				Commun. Chem. {\bf 1}, 96 (2018).



\bibitem{Mattos12}		T. G. Mattos, C. Mejia-Monasterio, R. Metzler, and G. Oshanin,
				First passages in bounded domains: When is the mean first passage time meaningful?,
				Phys. Rev. E {\bf 86}, 031143 (2012).

\bibitem{Godec16b}		A. Godec and R. Metzler,
				First passage time distribution in heterogeneity controlled kinetics: going beyond the mean first passage time,
				Sci. Rep. {\bf 6}, 20349 (2016). 

\bibitem{Grebenkov18b}		D. S. Grebenkov, R. Metzler, and G. Oshanin, 
				Towards a full quantitative description of single-molecule reaction 
				kinetics in biological cells, 
				Phys. Chem. Chem. Phys. {\bf 20}, 16393-16401 (2018).

\bibitem{Reva21}		M. Reva, D. A. DiGregorio, and D. S. Grebenkov, 
				A first-passage approach to diffusion-influenced reversible binding: 
				insights into nanoscale signaling at the presynapse, 
				Sci. Rep. {\bf 11}, 5377 (2021).


\bibitem{Yuste13}		S. B. Yuste, E. Abad, and K. Lindenberg,
				Exploration and trapping of mortal random walkers,
				Phys. Rev. Lett. {\bf 110}, 220603 (2013).

\bibitem{Meerson15}		B. Meerson and S. Redner,
				Mortality, Redundancy, and Diversity in Stochastic Search,
				Phys. Rev. Lett. {\bf 114}, 198101 (2015).

\bibitem{Grebenkov17f}		D. S. Grebenkov and J.-F. Rupprecht, 
				The escape problem for mortal walkers, 
				J. Chem. Phys. {\bf 146}, 084106 (2017).

\bibitem{Meerson19}		B. Meerson,
				Mortal Brownian motion: Three short stories,
				Int. J. Mod. Phys. B {\bf 33}, 1950172 (2019).




\bibitem{Bressloff22a}		P. C. Bressloff,
				Diffusion-mediated absorption by partially-reactive targets: 
				Brownian functionals and generalized propagators,
				J. Phys. A: Math. Theor. {\bf 55}, 205001 (2022).

\bibitem{Bressloff22b}		P. C. Bressloff,
				Narrow capture problem: an encounter-based approach to partially reactive targets,
				Phys. Rev. E {\bf 105}, 034141 (2022).




\bibitem{Grebenkov22b}		D. S. Grebenkov, 
				Depletion of Resources by a Population of Diffusing Species, 
				Phys. Rev. E {\bf 105}, 054402 (2022). 


\bibitem{Evans20}		M. R. Evans, S. N. Majumdar, and G. Schehr,
				Stochastic resetting and applications
				J. Phys. A: Math. Theor. {\bf 53}, 193001 (2020).

\bibitem{Bressloff22c}		P. C. Bressloff,
				Diffusion-mediated surface reactions and stochastic resetting,
				J. Phys. A: Math. Theor. {\bf 55}, 275002 (2022).

\bibitem{Benkhadaj22}		Z. Benkhadaj and D. S. Grebenkov, 
				Encounter-based approach to diffusion with resetting, 
				Phys. Rev. E {\bf 106}, 044121 (2022).


\bibitem{Grebenkov22a}		D. S. Grebenkov, 
				An encounter-based approach for restricted diffusion with a gradient drift, 
				J. Phys. A: Math. Theor. {\bf 55}, 045203 (2022).




\end{thebibliography}
\end{document}